\documentclass{article}

\usepackage{arxiv}

\usepackage[utf8]{inputenc} 
\usepackage[T1]{fontenc}    
\usepackage{hyperref}       
\usepackage{url}            
\usepackage{booktabs}       
\usepackage{nicefrac}       
\usepackage{microtype}      
\usepackage{lipsum}
\usepackage{graphicx}
\usepackage{latexsym}
\usepackage{multicol,multirow}
\usepackage{amsmath,amssymb,amsfonts}
\usepackage{mathrsfs}
\usepackage{amsthm}
\usepackage{apacite}
\usepackage{rotating}
\usepackage{appendix}
\usepackage[authoryear]{natbib}
\usepackage{ifpdf}
\usepackage[T1]{fontenc}
\usepackage{times}
\usepackage{sourcesanspro}
\usepackage{newtxmath}
\usepackage{textcomp}%
\usepackage{xcolor}%
\usepackage{subcaption}
\usepackage{siunitx}
\newcommand*{\vertbar}{\rule[0.1ex]{0.5pt}{2.5ex}}

\title{Sparse surface pressure-based reconstruction of the flow around a thick airfoil over a range of angles of attack}

\author{
 Quentin Bucquet \\
  EM2C, CentraleSupélec, \\
  Université Paris-Saclay, \\
  France\\
  \texttt{quentin.bucquet@centralesupelec.fr} \\
   \And
 Bérengère Podvin \\
  EM2C, CentraleSupélec, \\
  Université Paris-Saclay, \\
  France\\
  \And
 Caroline Braud \\
  LHEEA, CNRS, Centrale Nantes, \\
  Nantes-Université, \\
  France \\
    \And
 Emmanuel Guilmineau \\
  LHEEA, CNRS, Centrale Nantes, \\
  Nantes-Université, \\
  France \\
}

\begin{document}
	\maketitle
    \begin{abstract}
		
		We present an efficient neural-based approach to estimate the instantaneous flow field around an airfoil from limited surface pressure measurements. The model, denoted SNN-POD, relies on two independent shallow neural networks to predict the instantaneous flow over a wide range of angles of attack $[10^\circ,20^\circ]$. At all angles the global model correctly recovers the average characteristics of the flow from single-time sensor data, thus allowing combination with local, angle-dependent models. The method is applied to 2D URANS simulations of a thick airfoil at a Reynolds number of $Re=\num{4.5e6}$. The training set consists of snapshots obtained from a coarse sampling $(1-2^\circ)$ of the angle of attack range. A variance-based criterion is used to determine the number and positions of sensors. Tests are carried out for unseen snapshots at angles of attack within the set (sampled angles) as well as outside the set (interpolated angles). The maximum MSE error of attack for sampled and interpolated angles is respectively $2.9\%$ and $6.6\%$. This makes it possible to develop adaptive strategies to improve the estimation if necessary.
		
	\end{abstract}
		
    \keywords{Sparse Flow Reconstruction; Data-driven; Reduced-Order Modeling; Wind Energy}
		
			
			
	

	\section{Introduction}
	
	In the context of wind energy, accurate prediction of mean and instantaneous wind turbine aerodynamic loads is essential for ensuring the structural integrity and efficiency of wind turbines, particularly during gusts or dynamic stall. Originally introduced by \cite{Glauert1935}, the Blade Element Momentum (BEM) method has long been the workhorse for aerodynamic load prediction. Its appeal lies in its simplicity and computational efficiency, assuming stationary and uniform inflow. Over the decades, BEM has been refined with numerous corrections to account for three-dimensional effects such as tip losses \citep{prandtl1927vier}, root effects \citep{chaviaropoulos_investigating_2000}, yawed inflow \citep{wes-5-1-2020}, and dynamic inflow conditions \citep{etde_53347}. Based on comparisons with large-scale field test campaigns, \cite{potentier_analysis_2021} identified the most impactful corrections and demonstrated that, when supplemented with unsteady aerodynamic models, BEM can reliably capture mean load trends in most conditions. However capturing instantaneous loads remains an important challenge, as it requires knowledge of the surrounding flow field—information that is challenging to access in practical or in-situ settings where only sparse, low-resolution wall-pressure measurements are typically available.
	
	The field of sparse flow reconstruction, where a reliable estimator of the flow is built using only limited or noisy measurements, has been widely studied over the last decades in many engineering applications, ranging from medical diagnostics to atmospheric measurements. 
	In the medical field,  the spatially and temporally limited acquisition of the four-dimensional flow magnetic resonance imaging (4D MRI) is complemented by flow reconstruction techniques to determine blood flow during cerebral aneurysms \citep{zhang_multi-modality_2022}. Reconstruction of turbulent flows from restricted observations is also crucial for climatology \citep{schneider_earth_2017}, weather forecast \citep{kalnay_atmospheric_2002}, atmospheric \citep{kaimal_1994} and urban boundary layers studies \citep{blackman_non-linear_2016,lu_flow_2023}, where metrology techniques (meteorological masts, LIDAR, pressure balloons) give access to only a small fraction of the atmospherical volume of interest. 
	The same problem applies in wind energy when estimating the inflow wind or the downstream wake for both wind turbines and wind farms using high-frequency SCADA data on the turbines \citep{rott_wind_2020}.
	
	Several challenges for flow reconstruction need to be addressed. Firstly, the target flow field to estimate is often turbulent and therefore is characterized by high-dimensional data, which generally needs to be reduced down to a more compact representation. Secondly, physical and economic constraints can limit the number, position and acquisition rate of the sensors so that optimization strategies may be needed, \citep{hollenbeck24,bukowski2025sparse}. Thirdly, the mapping between the observations and the target field may be nonlinear, as for instance in the reconstruction of a velocity field from pressure measurements \citep{kn:graziani18}. Finally, an adequate compromise should be found between the cost and the accuracy of the estimation technique.
	
	
	Comprehensive reviews of sparse reconstruction methodologies can be found in \cite{callaham_robust_2019,nair_leveraging_2020,manohar_data-driven_2018}, but a few key principles are worth recalling here. The majority of approaches rely on Reduced-Order Models (ROMs) to project the high-dimensional flow dynamics onto a low-order manifold. The reconstruction task then reduces to estimating the reduced state, which can subsequently be mapped back to the full-order field. The projection of the high-dimensional dynamics of the complex flow field can be carried out using a linear decomposition basis 
	that can be data-agnostic, such as Fourier modes or wavelets, or data-driven, such as Proper Orthogonal Decomposition aka POD \citep{kn:lumleyPOD}, or Dynamic Mode Decomposition aka DMD \citep{schmid_dynamic_2010}. A comprehensive review of modal decompositions can be found in \cite{kn:taira17}. However, linear reconstructions may be limited in their ability to capture nonlinear flow features. This has motivated the development of more sophisticated methods such as sparse regression techniques to address these issues \citep{kn:loiseau18}. 
	
	Generally speaking, recent advances in machine learning have opened new perspectives for flow reconstruction : neural-network-based approaches can learn nonlinear mappings between sparse measurements and the full-order flow, although linear expansions can still be used to reduce the dimensionality of the data \citep{swischuk19}. Among the various neural network architectures that have attracted interest, autoencoders have shown promise for discovering nonlinear manifolds with improved reconstruction accuracy compared to linear subspaces \citep{lee_model_2019,xu_multi-level_2020,tan_flow_2023} and can provide causality insights \citep{fukamitaira25}. Graph neural networks (GNNs) have also been shown to be a versatile and powerful tool for estimation purposes, as was recently evidenced by \cite{duthe_graph_2023}, who developed a message-passing GNN to predict steady two-dimensional pressure and velocity fields from surface pressure distributions around airfoils of arbitrary shape. 
	
	Possible drawbacks of data-driven neural frameworks include their lack of interpretability and their disconnection from physics-based knowledge. These limitations are addressed by Physics-Informed Neural Networks (PINNs), introduced by \cite{raissi_physics-informed_2019}, which are trained with a loss that seeked to both minimize the discrepancy with the observations and to enforce physics conservation laws. Implementations included velocity reconstruction in a cylinder wake from sparse and incomplete velocity measurements \citep{xu_practical_2023}, bluff-body wake and Kolmogorov flow reconstruction with a Convolution Neural Network trained for various physics-based loss formulations \citep{mo_reconstructing_2025}, and reconstruction of three-dimensional turbulent wall-bounded flows from two-dimensional planar measurements with a physics-constrained variational autoencoder \citep{hora_physics-informed_2024}. However, while PINNs show great potential, they can induce high computational costs, and can struggle on complex flows, such as multi-scale dynamic systems \citep{li_physics-informed_2024}.
	
	In contrast, a low-cost alternative that has shown substantial promise is the shallow neural network model introduced by \cite{erichson_shallow_2020} (referred therein as "shallow decoder", an denoted as SNN by \citealp{carter_data-driven_2021}), which relies on a simple neural architecture to map sparse sensor inputs to high-dimensional flow fields. The shallow neural network model has demonstrated superior performance compared to POD-based linear reconstructions in test cases such as flow past a cylinder, sea-surface temperature dynamics, and synthetic turbulence \citep{erichson_shallow_2020}. Shallow neural networks have further been used as nonlinear refinements of linear state estimation \citep{carter_data-driven_2021} or as ROM-based models for reconstructing vorticity fields from vorticity- and surface stress-measuring sensors on the body of the flat plate at high angle of attack \citep{nair_leveraging_2020}. The framework of shallow neural networks has also been extended to taken into account temporal information. \cite{williams_sensing_2024} successfully tested a shallow recurrent decoder (SHRED) that relies on sensor past history for reconstruction and forecasting tasks. This approach can be complemented by a ROM (SHRED-ROM, \citealp{kutz_shallow_2024,tomasetto_reduced_2025}).
	
	In the present work, a hybrid approach blending POD and shallow neural networks is used to reconstruct the flow velocity in the wake of a thick airfoil using sparse, instantaneous surface pressure measurements. The dataset consists of unsteady two-dimensional URANS simulations at high Reynolds number obtained over various angles of attack. The model output or target is constituted by the instantaneous POD amplitudes of the flow velocity, which are split into a time-dependent and a time-independent parts, each of which is learnt independently by a shallow neural network.
	
	The remainder of this paper is structured as follows. Section \ref{sec:dataset_presentation} describes the URANS simulations from which the training and test datasets are constructed, with a focus on the surface pressure signal (input) and the streamwise velocity field (final output). The reconstruction procedure is detailed in Section \ref{sec:methodology}. Evaluation of the approach and of its robustness is carried out in Section \ref{sec:results}. A conclusion and discussion of 
	future developments are given in Section \ref{sec:conclusion}.
	
	\section{Dataset presentation}
	\label{sec:dataset_presentation}
	
	\subsection{Simulation set up}
	
	The present dataset consists in two-dimensional simulations of a wind turbine airfoil for angles of attack ranging from $10^\circ$ to $20^\circ$. 
	It is generated with the ISIS-CFD flow solver \citep{visonneau_2014}, developed by Centrale Nantes and CNRS and part of the FINE™/Marine computing suite, using an incompressible Unsteady Reynolds-Averaged Navier Stokes (URANS) approach. 
	The $k-\omega$ SST model is used for turbulence modeling. 
	A finite volume method spatially discretizes the transport equations on an unstructured mesh which is automatically prolongated (refined) or restricted (derefined) using Adaptive Mesh Refinement (AMR) to accurately capture local gradients without leading to excessive computational costs. 
	Flow variables are stored at the center of the arbitrary shaped cells. 
	Volume and surface integrals are computed using second-order approximations, whereas fluxes at the mesh faces 
	are reconstructed by linear extrapolation of the integrand from the neighboring cell centers. 
	A second-order backward difference scheme is used for time discretization. 
	A constant time step computed as $\Delta t U_{\infty} / c = 0.01$  convective time units, where $U_\infty$ is the inlet velocity and $c$ the blade chord, is applied for all the simulations, and snapshots are stored every 5 time steps. 
	
	The airfoil geometry is extracted from a 3D scan of a 2MW wind turbine blade at $80\%$ of the rotor diameter, featuring a chord length $c=1.25$ m and a thickness of $20\%$. 
	The chord-based Reynolds number is $R_e =\num{4.5e6}$, which reproduces the full-scale geometry in the CSTB climatic wind tunnel \citep{neunaber_wind_2022,braud_study_2024}. 
	A comparison of the the experimental and numerical global loads was carried out in \cite{neunaber_wind_2022} and showed a good agreement. 
	
	For each angle of attack, the simulation was run for at least $20000$ time steps, corresponding to approximately $200$ convective time units, and required approximately $300$ core hours of computational time.
	While AMR is used when running the simulations, the post-processing and training / testing of our flow reconstruction methodology are performed on 
	a fixed grid, taken as the mesh of the last iteration of the $14^\circ$ simulation, consisting of $N_s = \num{91912}$ grid points. 
	
	\subsection{Angle of attack discretization}
	\label{sec:AoA_discretization}
	
	The angles of attack (AoA) featured in the training set form a set $I_{train}=\{10^\circ, 12^\circ, 14^\circ, 15^\circ, 16^\circ, 18^\circ, 20^\circ\}$. 
	This range is selected to cover various flow regimes: cases with fully attached flows, transitional cases where the boundary layer starts to separate from the trailing edge (around $14^\circ$ in the simulations, $12^\circ$ in the experiments \citealp{braud_study_2024}), and cases where the boundary layer is separated, with the separation point moving upstream toward the leading edge. 
	This diversity of AoA reflects the natural variability of the angle of attack for a wind turbine due to changes in rotation speed and pitch in normal operating conditions. Additionally, the turbine and the airfoil are immersed in a complex turbulent atmospheric environment which induces strong local fluctuations of the angle of incidence due to wind gusts, wind ramps (i.e. rapid changes in incoming wind speed) or yaw misalignment, for instance. 
	The angle of attack increments were of $1^\circ$ in the range $[14^\circ,16^\circ]$, which corresponds to the onset
	of separation, and $2^\circ$ otherwise.
	This coarse discretization over the range of AoA was chosen in order to allow fast training and to evaluate the generalization capabilities of the model.
	
	\subsection{Wall pressure}
	
	This section  describes the pressure measurements  on the airfoil, a  subset  
	of which will constitute the model inputs. 
	
	\subsubsection{Wall pressure distribution}
	
\begin{figure}[h]
		\centering
		\begin{subfigure}{.48\textwidth}
			\centering
			\includegraphics[width=\linewidth]{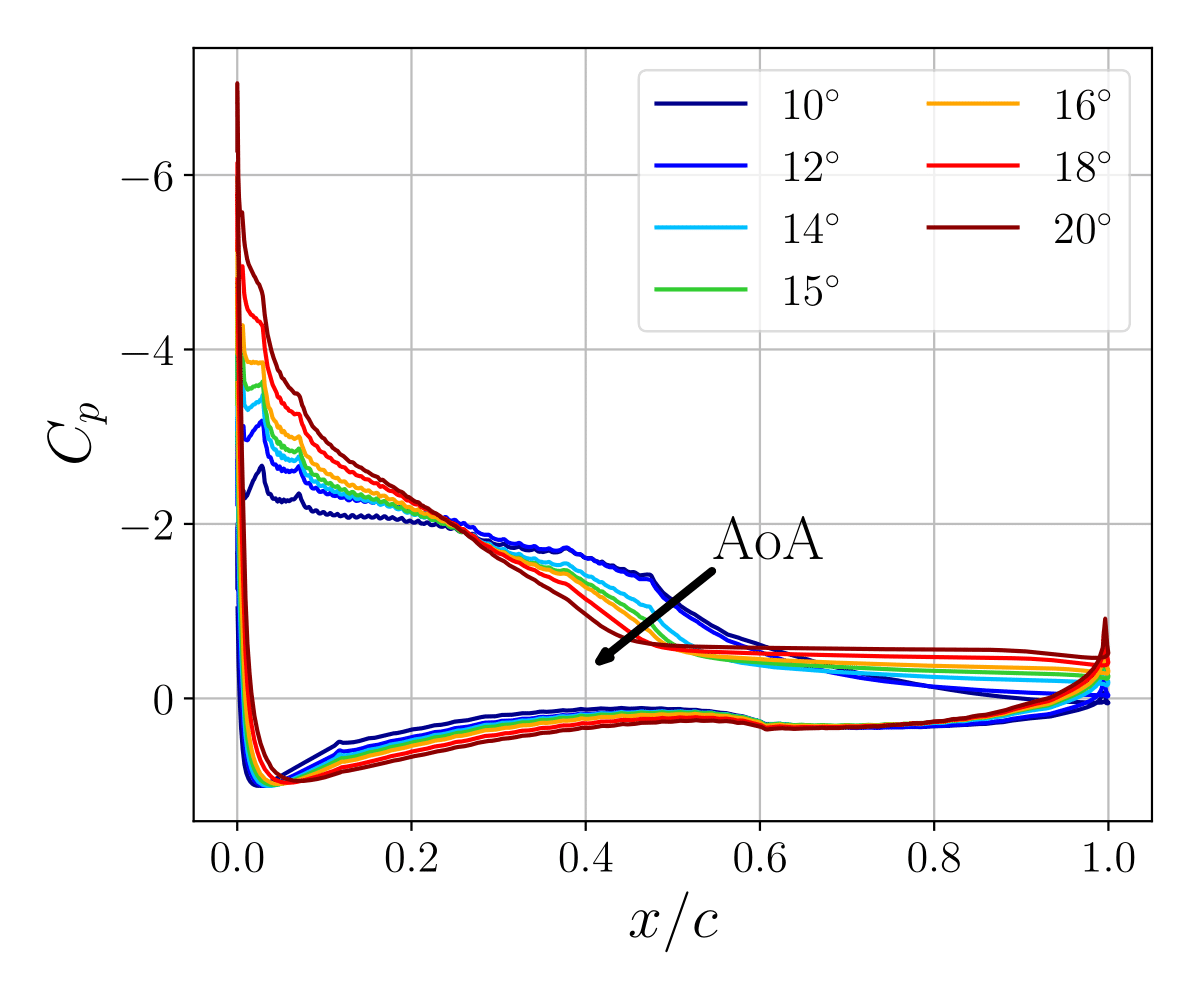}
			\caption{}
		\end{subfigure}%
		\begin{subfigure}{.48\textwidth}
			\centering
			\includegraphics[width=\linewidth]{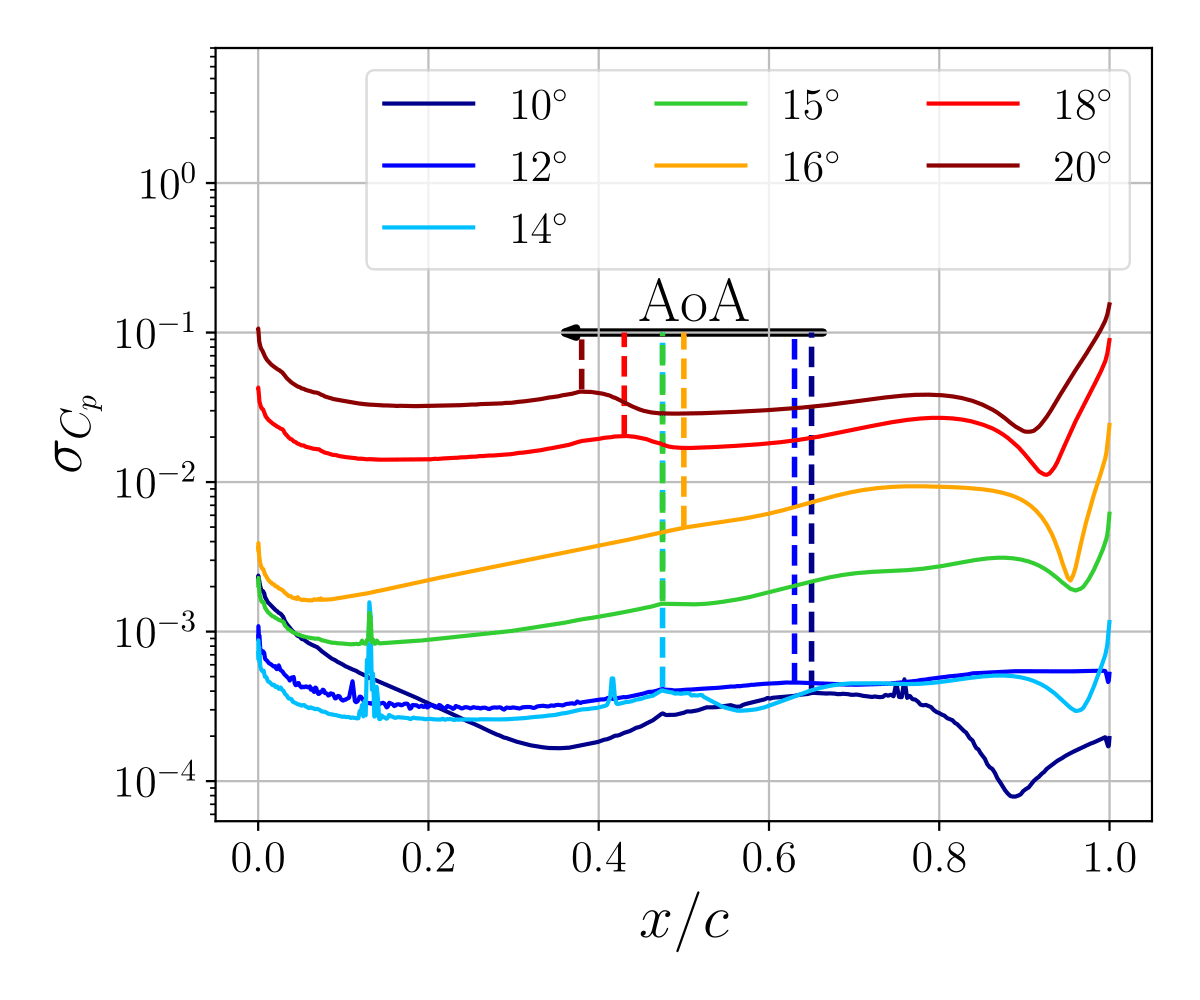}
			\caption{}
		\end{subfigure}
		\caption{a) Time-averaged pressure coefficient $C_p$ distribution around the airfoil  b) Standard deviation $\sigma_{C_p}$  of $C_p$ (suction side only) in chordwise direction for  angles of attack from $10^\circ$ to $20^\circ$. Vertical dashed lines correspond to the chordwise location of the Intermittent Separation Point (ISP), which is defined as the local maximum of $\sigma_{C_p}$ on the mid-chord pressure suction side}
		\label{fig:Cp_std_Cp}
	\end{figure}

	The distribution of the time-averaged wall pressure coefficient $C_p = \frac{\Delta p}{q_0}$, where $\Delta p$ is the differential between the wall pressure and the reference static pressure taken at freestream, and $q_0$ is the dynamic pressure, is shown in Fig. \ref{fig:Cp_std_Cp}(a) for AoA $\in I_{train}$. 
	For all angles of attack, the distributions exhibit similar characteristics : a strong suction peak at the leading edge that increases with AoA, a high suction region 
	in the first half of the blade, followed by a slow decrease of the suction down to the trailing edge. 
	It should be noted that some asperities in the profile are due to discontinuities in the geometry, which was directly obtained from a coarse 3D scan of a real-life blade.
	Beyond separation, for  AoA $\ge 14^\circ$, the progressive flattening of the curves with AoA near the trailing edge 
	indicates a shift of the flow separation point towards the leading edge. 
	This evolution is essentially similar to that described in \cite{neunaber_wind_2022} and \cite{braud_study_2024} for the 
	corresponding experimental configuration (however, in the experiment, separation was observed around AoA $\ge 12^\circ$ which can be attributed to slight differences in the ambient environment that is more noisy experimentally).  
	
	Figure \ref{fig:Cp_std_Cp}(b) shows the distribution of the standard deviation of the wall pressure coefficient, $\sigma_{C_p}$. 
	The level of pressure fluctuations remains very low for angles of attack where the flow remains attached or starts to separate ($10^\circ$, $12^\circ$, $14^\circ$). For higher angles of attack associated with separated flow regimes, $\sigma_{C_p}$ increases progressively. At high Reynolds number, this monotonic increase with the angle of the attack is consistent with experimental observations (\citealp{broerenbragg01}, B. Hanna, private communication). 
	For all angles of attack, a distinct Leading Edge Maximum, associated with fluctuations near the suction peak, is observed. 
	A local maximum of $\sigma_{C_p}$  was also consistently observed across all angles of attack in the chordwise region $x/c \in [0.3; 0.7]$. 
	It shifts gradually toward the leading edge as the angle of attack increases, with an approximately constant rate of displacement.
	The location of the fluctuation maximum was called the Intermittent Separation Point (ISP) in \cite{braud_study_2024} as it characterizes the region where intermittent separation point takes place.
	Another peak—referred to as the Trailing Edge Maximum—appears around $x/c \approx 0.8$, linked to the unsteady interaction between the separated shear layer and the airfoil surface in the trailing-edge region.
	These three regions of large fluctuations were also identified in the experiment \citep{braud_study_2024} for individual chords.
	However some differences were noted with the experiment: fluctuation levels associated with the ISP were found to be higher in the experiment, and the additional peak found at the trailing edge in the simulation above separation was not observed there.  
	These differences could be respectively explained by the 2D character of the simulation and the relatively low resolution of the experiment. 
	
	\subsubsection{Wall pressure fluctuations}
\begin{figure}
		\begin{tabular}{ccc}
			\centering
			\includegraphics[trim = 0 0 2.5cm 0, clip,  height=5cm]{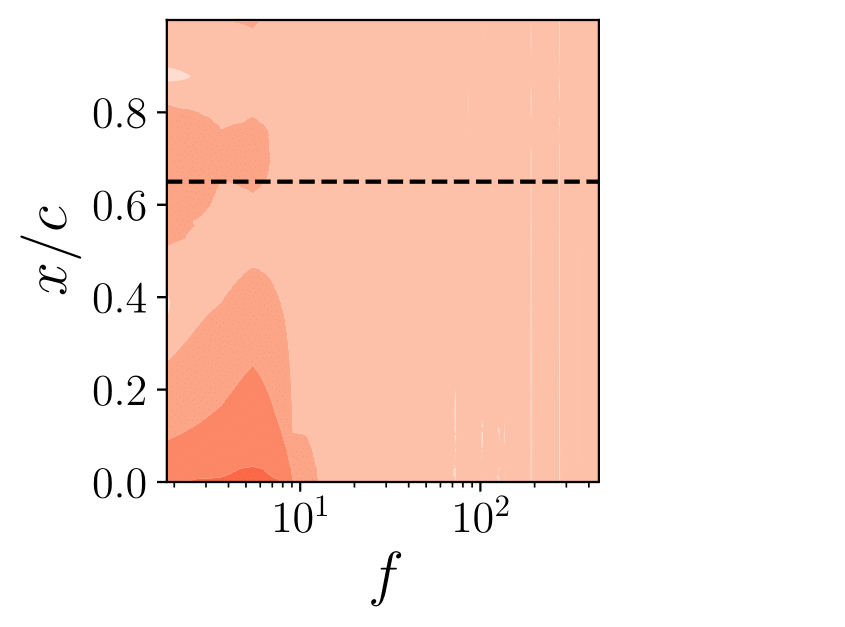} & \includegraphics[trim = 2cm 0 2.5cm 0, clip,  height=5cm]{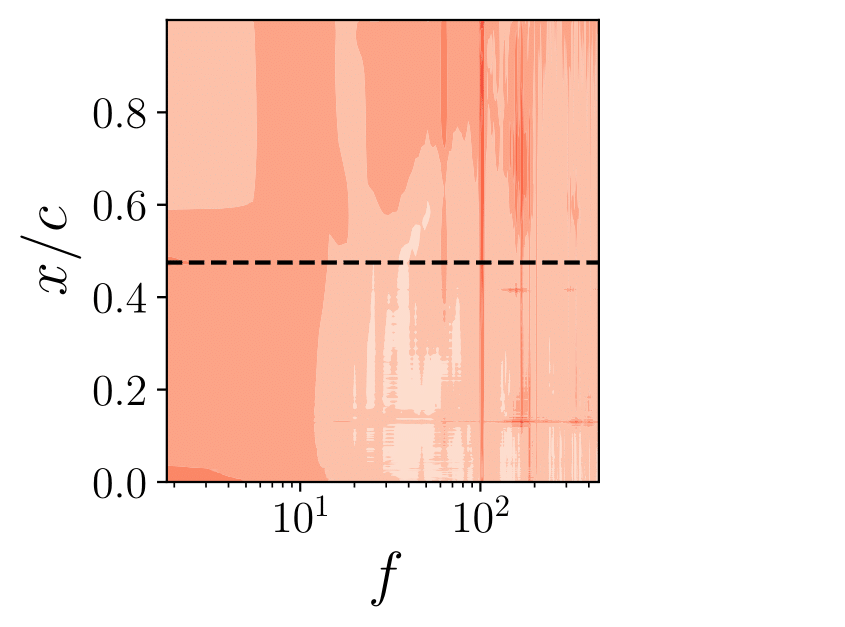} & \includegraphics[trim = 2cm 0 0 0, clip,  height=5cm]{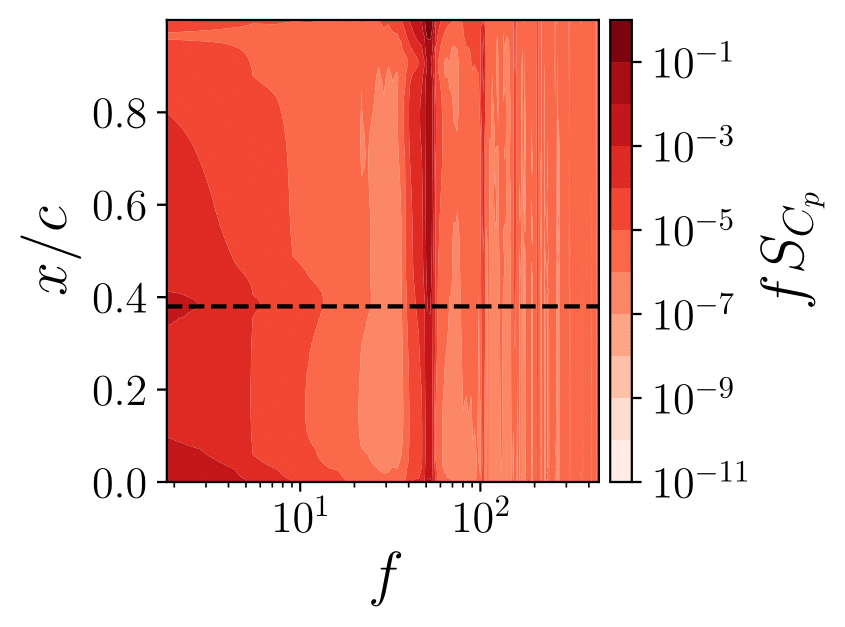} \\
			(a) $10^\circ$ & (b) $14^\circ$ & (c) $20^\circ$
			
		\end{tabular}
		\caption{Pre-multiplied time spectrograms of $C_p$ for AoA = $10^\circ$ (a), AoA = $14^\circ$ (b), AoA = $20^\circ$ (c). Black dashed lines corresponds to the location of the ISP}
		\label{fig:Cp_spectrograms}
	\end{figure}

	In this section and in the remainder of the work, the frequency auto-spectrum of the variable $x$ is identified as $S_x(f)$, with $f$ denoting the frequency.
	Fig. \ref{fig:Cp_spectrograms} shows the pre-multiplied frequency spectrograms of the pressure signal $f S_{C_p}$ for the suction side of the blade at three different angles of attack corresponding to attached flow ($10^\circ$), the onset of separation ($14^\circ$) and full separation ($20^\circ$).
	As expected from figure \ref{fig:Cp_std_Cp}(b), the energy levels increase with AoA.
	Below separation, low-frequency components ($f \leq 10$ Hz), typically associated with the strong suction peak dynamics, dominate near the leading edge. 
	At the onset of separation, a sharp frequency band centered at 100 Hz emerges from the spectrum. The  frequency band width increases and its center 
	increases with the angle of attack, as it reaches  around 50 Hz at $20^\circ$.
	A Strouhal number based on the freestream velocity $U_{\infty}$ and the shear layer height $h$ determined from the chord separation point and AoA yields a nearly constant value of $S_t = \frac{f h}{U_\infty} \sim 0.26$, which is consistent with the literature on 2D bluff bodies \citep{williamson96}. 
	
	\subsection{Wake characterization}
	\label{sec:wake_characterization}
	
	This section describes the main features of the velocity field, referred to hereafter as the "ground truth". 
	Throughout the paper, the focus of the reconstruction  will be  the instantaneous streamwise velocity component, although a similar appproach could be carried out for the cross-stream component. 
	Time-averaged then turbulent features are presented below.

	\subsubsection{Time-averaged wake}
	
	To characterize the mean features of the wake velocity deficit, defined as $\Delta u(x,y) = (U_\infty - \overline{u}(x,y))$, where
	$\overline{.}$ represents a time average, the following two quantities are introduced:  
	\begin{itemize}
		\item The wake width $\delta$ is defined as the spanwise distance at which the time-averaged velocity deficit ${\Delta u}(x,y)$ reaches a fraction $\theta$  of the local maximum velocity deficit, such that ${\Delta u}(x,y=\delta) = \theta {\Delta u}_{\text{max}}(x)$. Although a common criterion in wind turbine wake modeling is to use $\theta=0.1$, a more conservative threshold $\theta = 0.5$ was used as it was found to provide a more robust estimate with respect to small-scale perturbations in the surrounding flow \citep{duda_wake_2021}. 
		\item The wake deflection angle $\alpha$ is defined as the angle between the line corresponding to the locations of the maximum velocity deficit and the chord. 
		It is computed as $\alpha = \arctan(\frac{y_{\text{max}}-y_0}{x_{\text{max}}-x_0})$, where ($x_{\text{max}},y_{\text{max}}$) are the coordinates of the maximum velocity deficit and ($x_{0},y_{0}$) the coordinates of the trailing edge.
	\end{itemize}
	
	\begin{figure}[h]
		\centering
		\begin{subfigure}{.48\textwidth}
			\centering
			\includegraphics[width=\linewidth]{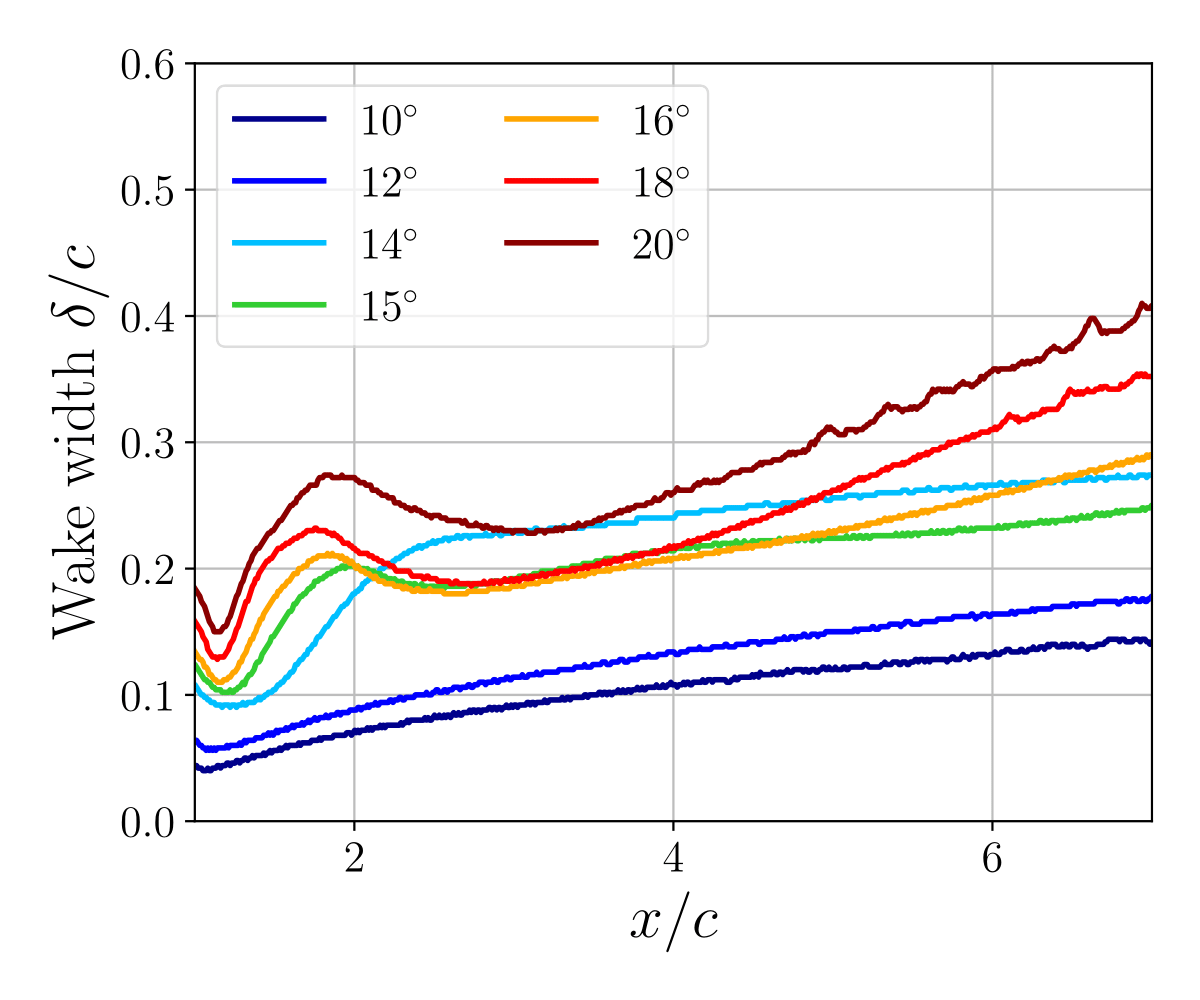}
			\caption{}
		\end{subfigure}%
		\begin{subfigure}{.48\textwidth}
			\centering
			\includegraphics[width=\linewidth]{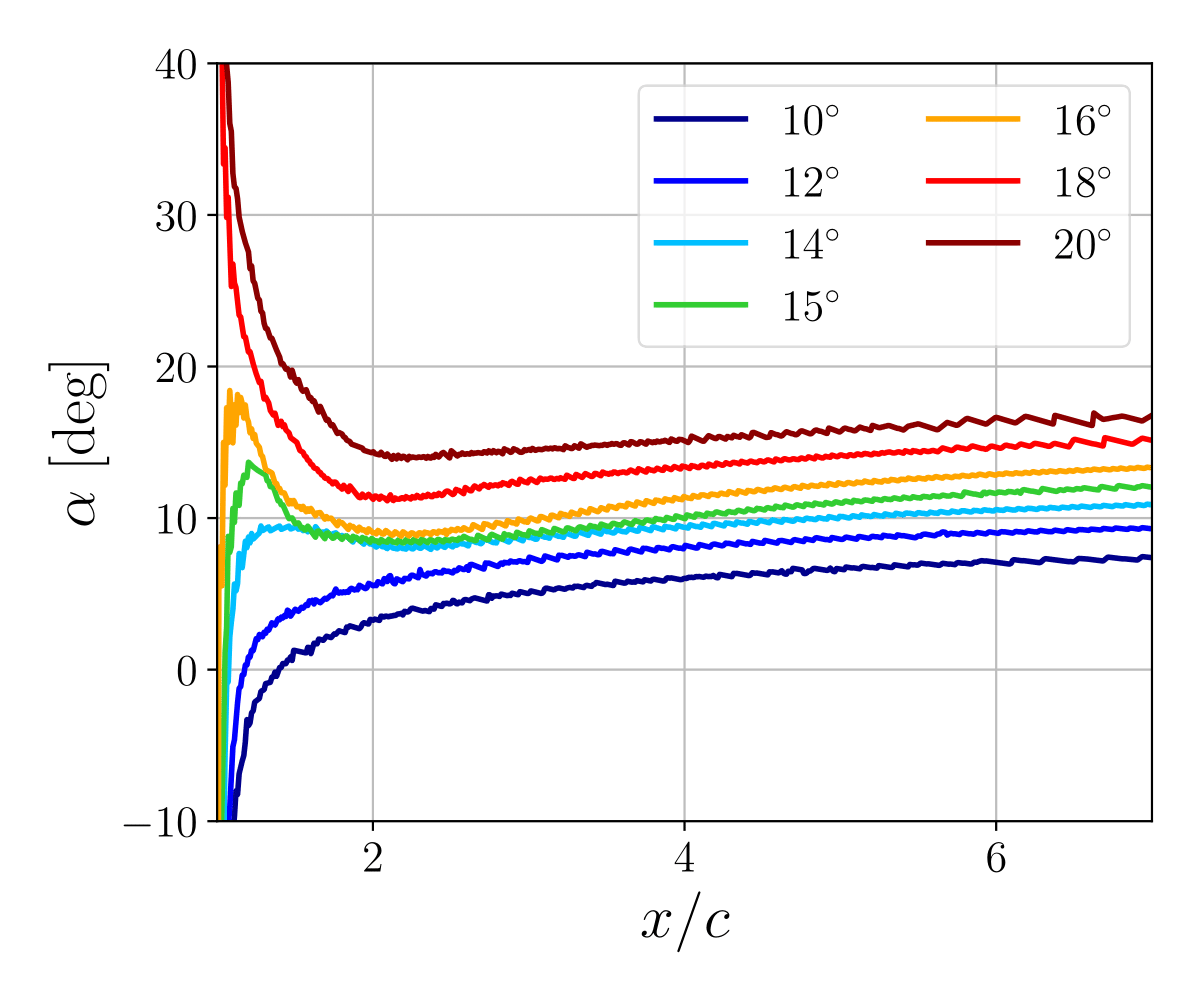}
			\caption{}
		\end{subfigure}
		\caption{Streamwise evolution of the time-averaged wake width $\delta(x)/c$ (a) and streamwise evolution of the time-averaged wake deflection angle $\alpha$ (b). The profiles are shown for $x/c \in [1;7]$, where $x/c = 1$ corresponds to the streamwise location of the trailing edge}
		\label{fig:streamwise_wake}
	\end{figure}
	
	Fig. \ref{fig:streamwise_wake} depicts the streamwise evolution of  the wake width and the wake deflection angle for all angles of attack. 
	Concerning the wake width, a general trend observed is that both $\delta$ and $d \delta / d x$ increase with the angle of attack. 
	For separated flow regimes, a specific trailing edge effect can be identified in the near wake ($x/c \in [1;2]$), as the width $\delta(x)$ first goes through a local minimum then a local maximum, and $d \delta / d x$ jumps to higher values. 
	At each angle of attack, the far wake evolution in the streamwise direction ($x>4c$) can be characterized by a nearly constant growth rate, which increases with the angle of attack. 
	Figure \ref{fig:streamwise_wake} also shows that the deflection angle  $\alpha$ increases with the angle of attack, with large variations in the near wake, and a quasi-linear increase in the far wake (it should be noted that the small oscillations observed there are an artefact of the computational mesh).
	The existence of a monotonic relationship thus makes it possible to infer the angle of attack from the deflection angle $\alpha$.
	The noisy behavior observed in the far wake profiles of both $\delta$ and $\alpha$ can be attributed to the mesh coarsening away from the airfoil.
	
	\subsubsection{Wake flow fluctuations}
	
\begin{figure}
		\begin{tabular}{ccc}
			\centering
			\includegraphics[trim = 0.3cm 0 2.3cm 0, clip,  height=7.5cm]{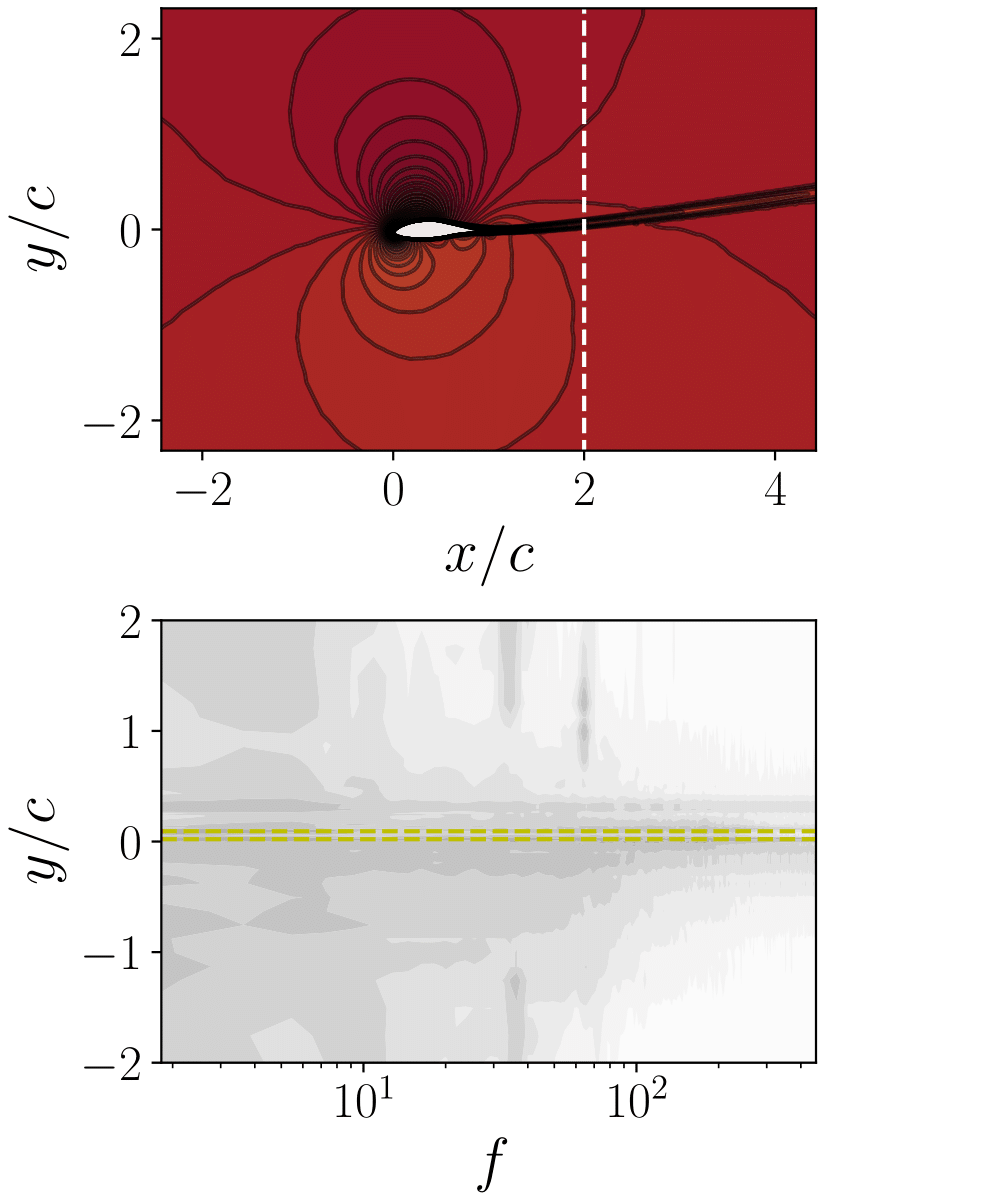} & \includegraphics[trim = 2cm 0 2.3cm 0, clip,  height=7.5cm]{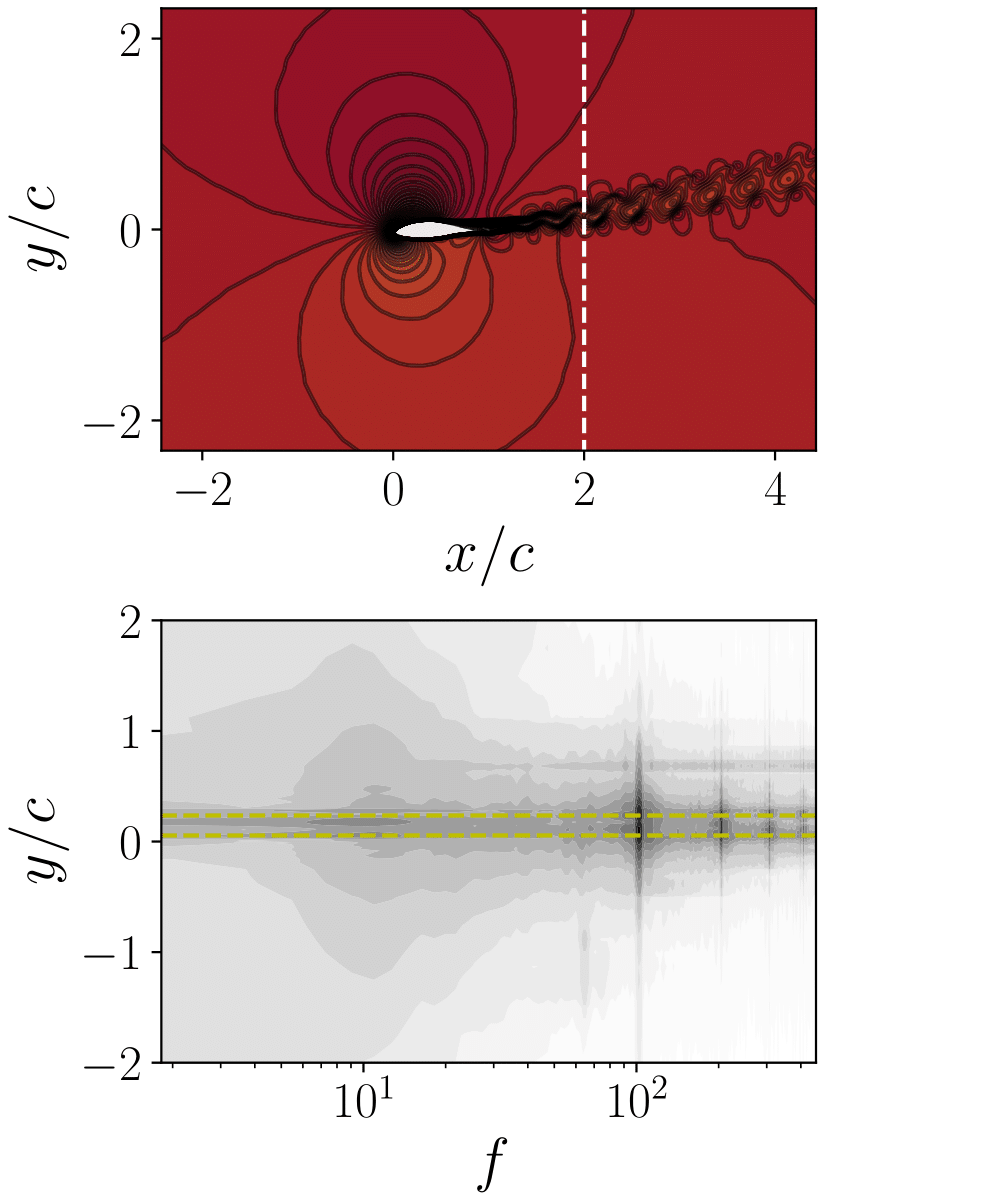} & \includegraphics[trim = 2cm 0 0 0, clip,  height=7.5cm]{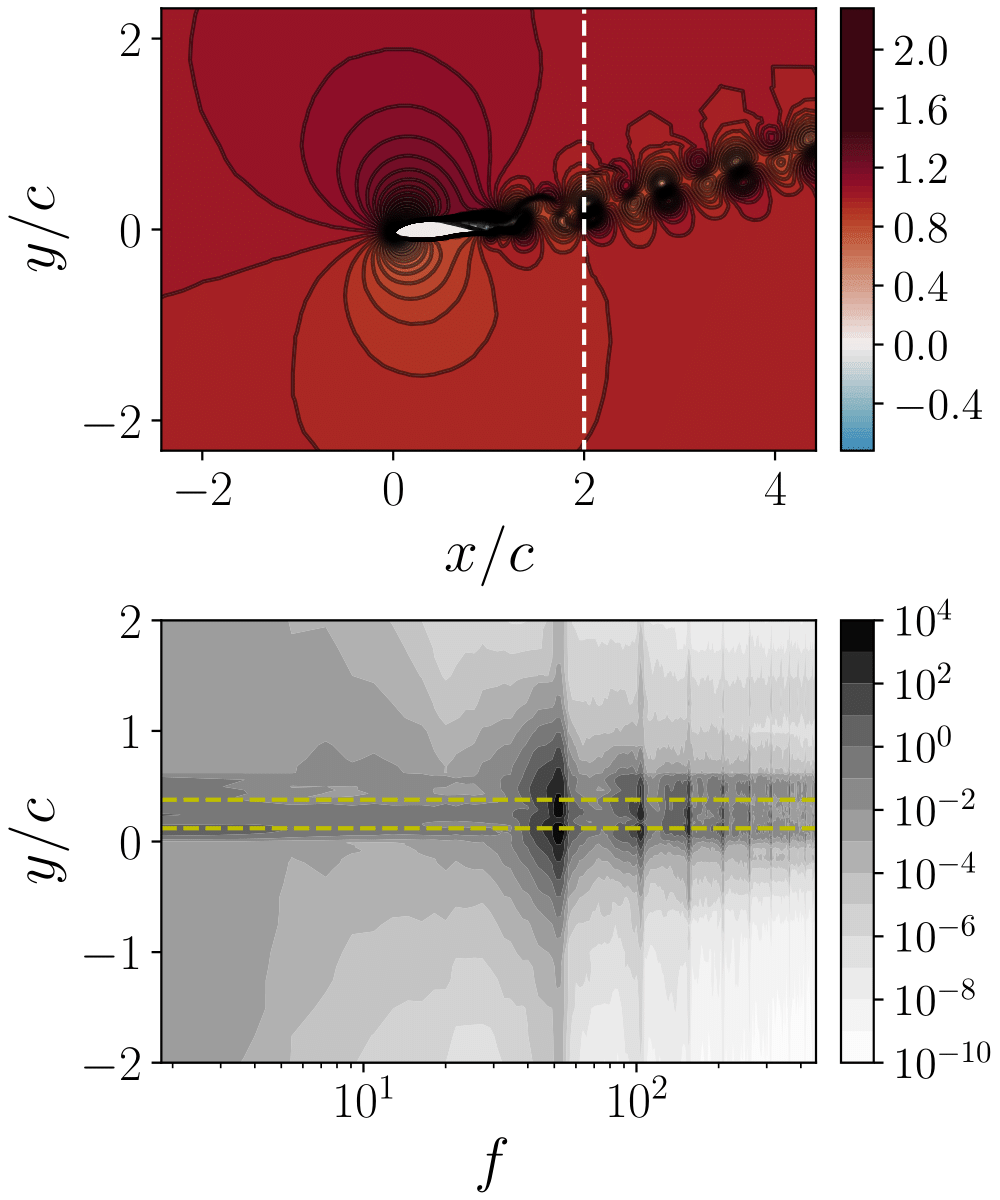} \\
			(a) $10^\circ$ & (b) $14^\circ$ & (c) $20^\circ$
			
		\end{tabular}
		\caption{Snapshots of the normalized instantaneous streamwise velocity field $u/U_\infty$ (top) and spectrograms of the pre-multiplied velocity spectra $f S_u$ for the cases AoA $=10^\circ$ (a), AoA $= 14^\circ$ (b), and AoA $= 20^\circ$ (c). The vertical white dashed lines in the snapshots denote the $x/c=2$ section where the spectrograms are computed. The horizontal yellow dashed lines in the spectrograms indicate the wake width}
		\label{fig:U_snapshots}
	\end{figure}

	Since the reconstruction method aims to recover instantaneous velocity fields, attention is now turned to the time-dependent features of the wake. 
	Snapshots of the instantaneous streamwise velocity field at an arbitrary time $t=5$s are shown in the top row of Fig. \ref{fig:U_snapshots} at the same angles of attack as those in Fig. \ref{fig:Cp_spectrograms}. It is recalled that all fields shown in the paper are rotated by placing the airfoil at a $0^\circ$ incidence and aligning the flow at the correct angle with respect to the AoA). 
	The vertical white dashed line at $x/c=2$ (near wake) corresponds to the section where pre-multiplied velocity spectra are extracted and shown on the bottom row of Fig. \ref{fig:U_snapshots}.
	
	As observed for the pressure field, the total energy of the fluctuations increase with the angle of attack. 
	The dominant frequencies observed in the pressure spectra in Fig. \ref{fig:Cp_spectrograms} are also present in the pre-multiplied velocity spectra $f S_u$
	(100 and 50 Hz for AoA = $14^\circ$ and $20^\circ$) and can be found to correspond to vortex shedding dynamics.  
	However several differences between pressure and velocity spectra can be noted, thus reflecting the nonlinear relationship between the pressure and velocity that the model needs to handle.\\
	
	
	In this section, we have characterized the inputs and outputs of the model to be described later, namely the surface pressure and the streamwise velocity field. Both time-averaged and time-dependent flow properties, such as the mean wake and its spectral signature, will be used to assess the model's performance.
	
	\section{Methodology}
	
	We now detail the estimation method for reconstructing the streamwise velocity field from a limited number of pressure measurements.
	We first describe how the sensor number and location are defined.
	
	\label{sec:methodology}
	
	\subsection{Sensor placement and number}
	\label{sec:sensors_placement_description}
	
	Owing to cost and practical constraints, the number of sensors needs to be kept low in real-life applications. 
	The sensor location is then critical because the information content is not uniformly distributed over the blade surface : regions such as the leading edge, trailing edge, or the intermittent separation zone carry substantially more information about the surrounding flow. 
	
	Much effort has been devoted to sensor placement optimization (see for instance \citealp{manohar_data-driven_2018,hollenbeck24, bukowski2025sparse, carter_data-driven_2021}).
	The approach of \cite{manohar_data-driven_2018} was followed here. 
	It consists in placing the sensors in order to maximize their contribution to the variance. 
	However, the variations of the variance with the angle of attack need to be taken into account, so that the information content associated with low fluctuations at the onset of separation does not get drowned out by the energetic fluctuations at higher angles of attack. 
	
	\begin{center}
		\begin{table}[h!]
			\centering
			\begin{tabular}{ l }
				\hline
				\textbf{Algorithm} Maximum variance approach \\
				\hline
				1: cluster blade points in $n_{c}$ clusters based their positions. \\
				2: define a variance threshold $\sigma^2_{\text{thres}}$ to be represented by sensors. \\ 
				3: \textbf{for} each $\text{AoA}_i$ \textbf{do} \\
				4: \hspace{0.5cm} compute the total variance of $C_p$ for each cluster \\
				5: \hspace{0.5cm} keep first $p_i$ most energetic clusters \\
				6: \hspace{0.5cm} choose locations of variance maximum within these $p_i$ clusters as sensor locations \\
				7: \hspace{0.5cm} \textbf{if} selected locations are too close \textbf{do} \\
				8: \hspace{1cm} move point of lesser variance up to a minimum distance $d_{min}$ \\
				9: \hspace{0.5cm} \textbf{end if} \\
				10: \textbf{end for} \\
				11: concatenate all selected sensors \\
				12: \textbf{if} concatenated sensors are too close \textbf{do} \\
				13: \hspace{0.5cm} keep point of higher variance, remove the other\\
				14: \textbf{end if} \\
				\hline
			\end{tabular}
			\caption{Methodology for selecting the wall pressure sensors based on maximizing the represented variance}
			\label{tab:sensor_location_algo}
		\end{table}
	\end{center}
	
	The first step consists in clustering blade surface points into $n_c$ clusters based on their positions.
	The k-means clustering method is used \citep{mcqueen67} and the number of clusters $n_c=25$ is determined using the elbow method \citep{thorndike1953belongs}.
	In the second step the optimal sensor locations are determined for each angle of attack : the wall-pressure variance is computed within each cluster, and the clusters are ranked according to their relative contribution to the total variance at this angle of attack.
	Given a variance threshold $\sigma^2_{\text{thres}}$ (expressed as a percentage of the total variance) to be captured, the minimal number of clusters $p_i$ necessary to capture this fraction is determined, and the corresponding clusters selected.
	In each of these $p_i$ clusters, one point only is selected (the point of maximum variance) as sensor location, with the assumption that the fluctuations of the point are representative of the fluctuations of the cluster.
	If two candidate sensors are found to be separated by less than $d_{min}$, the one associated with the highest variance is kept in place, whereas the other is moved within its cluster at a distance $d_{min}$.
	The minimum distance $d_{min}$ is taken to be $0.01 c$ and roughly corresponds to the pressure sensor spacing of \cite{neunaber_wind_2022}'s experiment. 
	The third step consists in concatenating the lists of  sensor positions obtained for the total number of angles of attack $N_{\text{AoA}}$.  
	At this last stage, if two sensors are found be at a distance less than $d_{min}$ from each other, only the sensor associated with the highest variance - regardless of the angle of attack - is retained. 
	The procedure yields a total number of $p \le p_i N_{\text{AoA}}$ sensors.
	The sensor selection process is summarized in table \ref{tab:sensor_location_algo}.
	
	\begin{figure}[h]
		\begin{subfigure}{.245\textwidth}
			\centering
			\includegraphics[width=\linewidth]{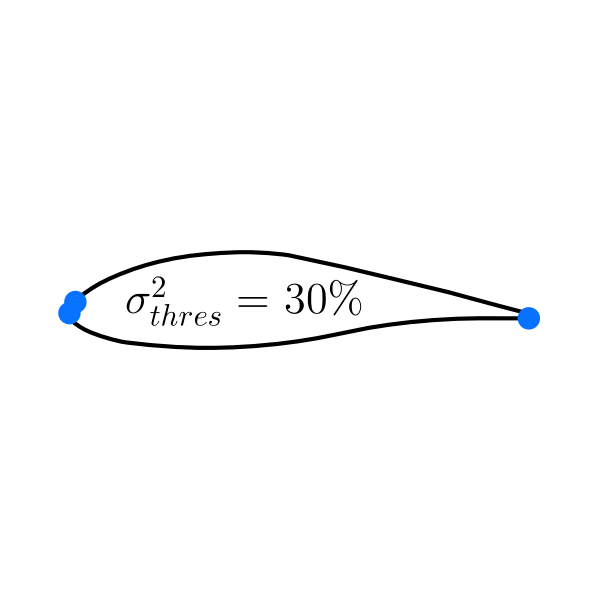}
		\end{subfigure}%
		\begin{subfigure}{.245\textwidth}
			\centering
			\includegraphics[width=\linewidth]{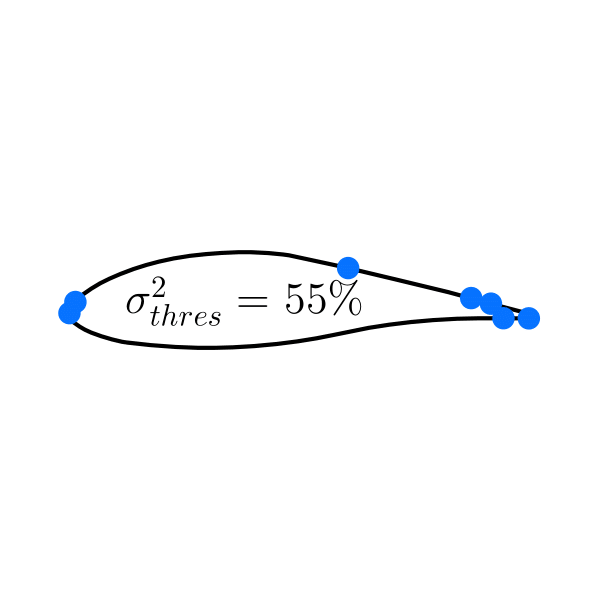}
		\end{subfigure}%
		\begin{subfigure}{.245\textwidth}
			\centering
			\includegraphics[width=\linewidth]{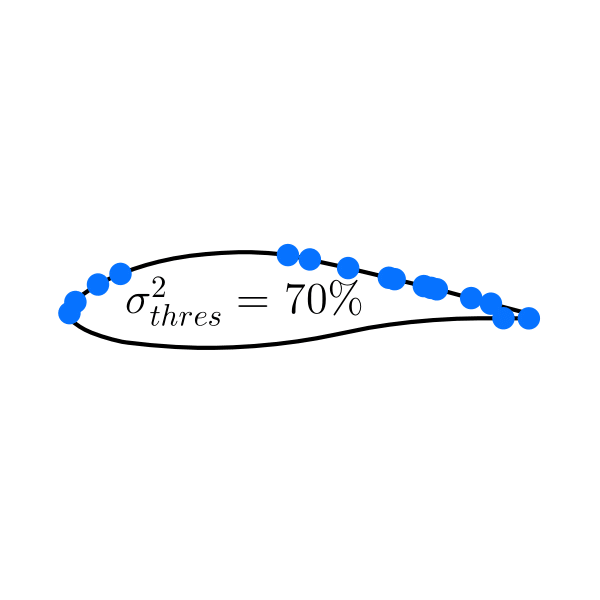}
		\end{subfigure}%
		\begin{subfigure}{.245\textwidth}
			\centering
			\includegraphics[width=\linewidth]{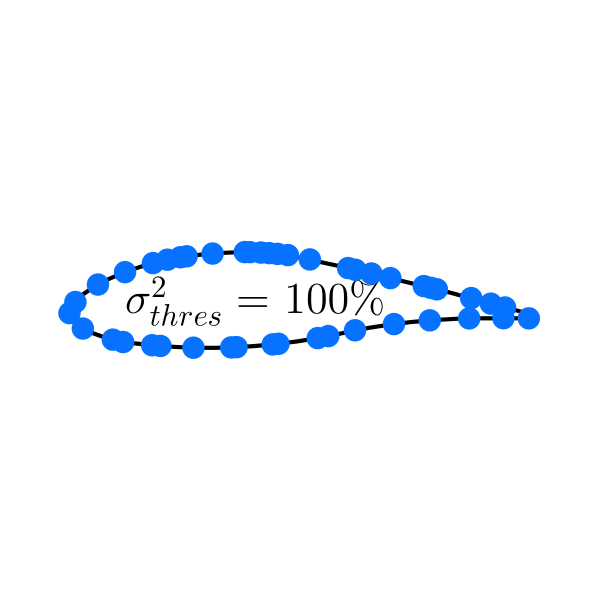}
		\end{subfigure}%
		\caption{Sensor layouts selected with the variance-maximization strategy for different threshold levels}
		\label{fig:sensors_loc}
	\end{figure}
	
	Table \ref{tab:sensor_nb_variance} reports the correspondence between the target variance threshold $\sigma^2_{\text{thres}}$ to be captured and the resulting number 
	of sensors $p$ required. 
	The relationship between the variance threshold and the number of selected sensors is strongly nonlinear. 
	At low threshold levels $(\le 30 \%)$, the number of sensors remains nearly constant or increases only marginally. 
	This can be attributed to the dominance of the Leading Edge Maximum and Trailing Edge Maximum, whose strong fluctuations alone account for up to $\sigma^2_{\text{thres}} = 30\%$ 
	of the total variance. 
	As the threshold increases, additional sensor positions are created — first in the leading and trailing edge regions, then  in the separation region, associated with the ISP, and 
	finally covering the surface of the airfoil, as shown in Fig. \ref{fig:sensors_loc}. 
	Based on the trade-off between reconstruction accuracy and the number of sensors, an optimal number of $p=7$ sensors (hence representing $55\%$ of wall pressure variance) 
	was  identified and will be used for most of the results presented below.
	This choice will be further detailed in Sect. \ref{sec:sensor_location_sensitivity}. 
	
	\begin{table}[h]
		\centering
		\begin{tabular}{ccccccccccc}
			\hline
			\hline
			$\sigma^2_{\text{thres}}$ (\%) & 10 & 20 & 30 & 40 & 50 & 60 & 70 & 80 & 90 & 100 \\
			\hline
			$p$ & 2 & 2 & 3 & 4 & 6 & 10 & 16 & 23 & 31 & 44 \\
		\end{tabular}
		\caption{Number of sensors $p$  as a function of the  variance fraction $\sigma^2_{\text{thres}}$} 
		\label{tab:sensor_nb_variance}
	\end{table}

	\subsection{The SNN-POD model}
	
	The SNN-POD model combines Proper Orthogonal Decomposition and the shallow decoder framework of \cite{erichson_shallow_2020} to reconstruct  
	the full velocity field from sparse wall pressure measurements taken over a range of angles of attack. 
	The model is trained with snapshots corresponding to the set $I_{train}$ of angles of attack of size $N_{\text{AoA}}=7$.
	For each angle of attack, a sequence of $N_t$ snapshots is selected to constitute the training set,
	which therefore consists of $N_t \times N_{\text{AoA}} $ input/ouput couples, where each input is a pressure measurement vector of size $p$, and each output is constituted by the snapshot of the velocity field of size $N_s$ with only one velocity component (streamwise).  
	Since the $m$-th snapshot, where $1 \le m \le N_t N_{\text{AoA}}$, can be associated to  a time $t_k$ and an angle of attack $\text{AoA}_i$,
	it will be referred to indifferently with the index $m$ or the tuple $(k,i)$. 
	The acquisition time resolution between snapshots was taken to be 0.05 convective time units (about 1.1 ms). 
	We used $N_{t} = 150$ snapshots, so that the length of the training set is about $0.16$s, which corresponds to $8-16$ vortex shedding cycles (defined using the vortex shedding frequencies observed in Fig. \ref{fig:U_snapshots}), depending on the angle of attack. 
	
\begin{figure}[h]
		\centering
		\includegraphics[width=0.85\linewidth]{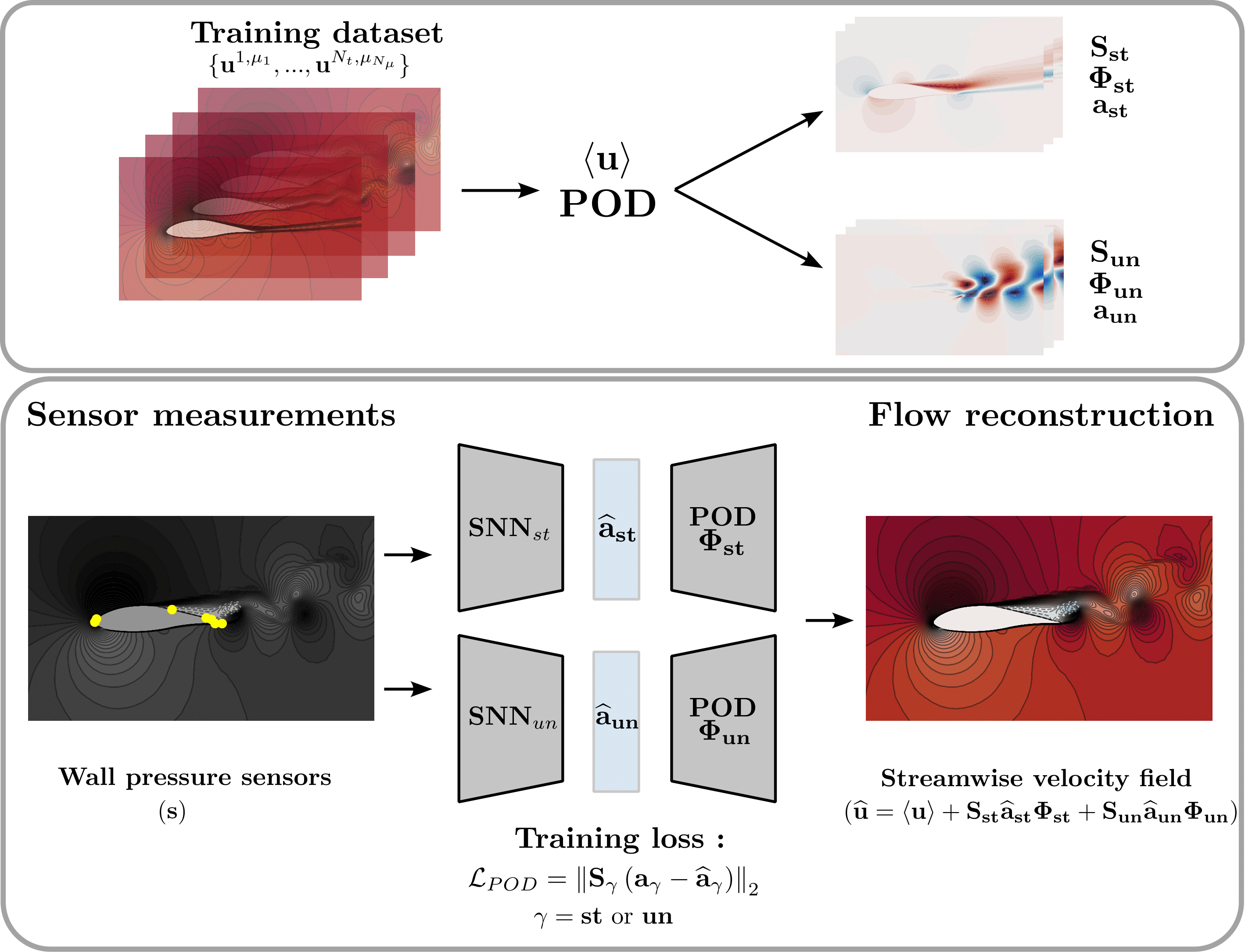}
		\caption{Schematic representation of the \textbf{SNN-POD} reconstruction framework. Steady (denoted with the $st$ subscript) and unsteady (denoted with the $un$ subscript) modes resulting from the Proper Orthogonal Decomposition of the training dataset are treated separately by the \textbf{SNN-POD}. Two Shallow Neural Networks (SNN) learn the mapping between the sparse wall pressure measurements and the steady and unsteady POD coefficients using  a energy-weighted loss. The high-dimensional flow field is then recovered from the POD training basis $\mathbf{\Phi}$ and estimated model amplitudes}
		\label{fig:model_scheme}
	\end{figure}

	
	The approach can be summarized as follows (see Fig. \ref{fig:model_scheme}) : POD is first used to compress information to a low-order manifold, a dimensionality reduction process which is presented in Sect. \ref{sec:dimension_reduction}. 
	A model based on shallow neural networks is then trained to learn the nonlinear mapping between the pressure measurements and the latent subspace (POD velocity amplitudes), the details of which are given in Sect. \ref{sec:SNN_architecture}. 
	The instantaneous flow field is then reconstructed from the temporal amplitudes estimated by the model and the spatial statistics of the training set.  
	
	\subsubsection{Dimensionality reduction}
	\label{sec:dimension_reduction}

	Due to the high dimension of the full-order state  to be reconstructed (see Sec. \ref{sec:dataset_presentation}), each velocity field in the training set 
	is first projected onto a low-dimensional linear subspace using Proper Orthogonal Decomposition \citep{kn:lumleyPOD}. 
	Since $N_t N_{\text{AoA}} < N_s$  the method of snapshots \citep{kn:siro87} is applied
	to the velocity fluctuation defined with respect to a mean $\langle u \rangle$, where $\langle . \rangle$ refers to an average taken over all times and angles of attack.
	The method relies on the construction of a snapshot matrix $\mathbf{X}$: 
	
	\[
	\mathbf{X} = 
	\left[
	\begin{array}{cccc}
		\vertbar & \vertbar &        & \vertbar \\
		\mathbf{u}^{1}    & \mathbf{u}^{2}    & \ldots & \mathbf{u}^{N_t N_{\text{AoA}}}    \\
		\vertbar & \vertbar &        & \vertbar 
	\end{array}
	\right]
	\]
	where  each column of $\mathbf{X} \in \mathbb{R}^{N_s \times N_t N_{\text{AoA}}}$ represents a velocity snapshot.
	It is to be noted that $\mathbf{u}^{1}$ and $u(\mathbf{x},t_1, \text{AoA}_1)$ will be considered as equivalent notations.

	Singular value decomposition (SVD) of the snapshot matrix $\mathbf{X}$ yields :
	$$
	\mathbf{X} = \mathbf{\Phi} \mathbf{S} \mathbf{A}^T
	$$
	where
	$$
	\mathbf{\Phi} = \left[
	\begin{array}{ccc}
		\mathbf{\phi}_{1}   & \ldots & \mathbf{\phi}_{N_t N_{\text{AoA}}}    \\
	\end{array}
	\right] \in \mathbb{R}^{N_s \times N_t N_{\text{AoA}}},
	\quad \quad \quad
	\mathbf{A} = \left[
	\begin{array}{ccc}
		\mathbf{a}_{1}   & \ldots & \mathbf{a}_{N_t N_{\text{AoA}}}    \\
	\end{array}
	\right] \in \mathbb{R}^{N_t N_{\text{AoA}} \times N_t N_{\text{AoA}}},
	$$
	$$
	\mathbf{S} = \text{diag}(\lambda_1^{1/2},\dots,\lambda_{N_t N_{\text{AoA}}}^{1/2}) \in \mathbb{R}^{N_t N_{\text{AoA}} \times N_t N_{\text{AoA}} },
	\quad \quad \quad
	$$
	
	with $\mathbf{S}$ the diagonal matrix of singular values ranked by magnitude, $\mathbf{\Phi}$ the matrix of linear POD modes and $\mathbf{A}$ 
	the matrix of POD time coefficients. 
	Each  POD mode $\phi_n (\mathbf{x})$, $n=1, \ldots N_t N_{\text{AoA}},$ satisfies the orthonormality condition as well each amplitude vector $\mathbf{a}_{n}$. 
	The value $\lambda_n$ represents the contribution of the $n$-th mode to the variance of the velocity field.  
	Each streamwise velocity field of the training set  can thus be expressed at any angle of attack $\text{AoA}_i$ with 
	a unique POD basis: 
	\begin{equation}
		\label{eq:POD_approx}
		u(\mathbf{x},t_k,\text{AoA}_i) = \langle u(\mathbf{x}) \rangle + u'(\mathbf{x},t_k,\text{AoA}_i) \mbox{ where }
		u'(\mathbf{x},t_k,\text{AoA}_i) = \sum_{n=1}^{N_t N_{\text{AoA}}} \lambda_n^{1/2} a_n^{k,i} \phi_n (\mathbf{x}) 
	\end{equation}
	
\begin{figure}
		 \begin{tabular}{ccc}
			     \centering
			     \includegraphics[trim = 0.1cm 0 2.7cm 0, clip,  height=4.6cm]{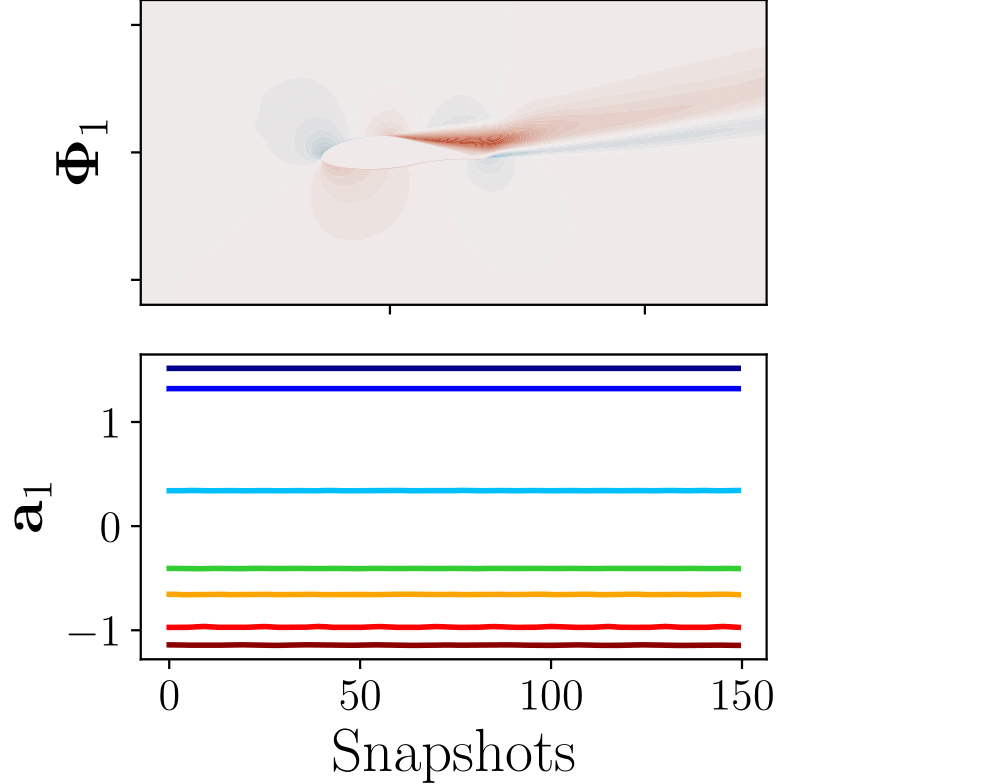} & \includegraphics[trim = 0.1cm 0 2.7cm 0, clip,  height=4.6cm]{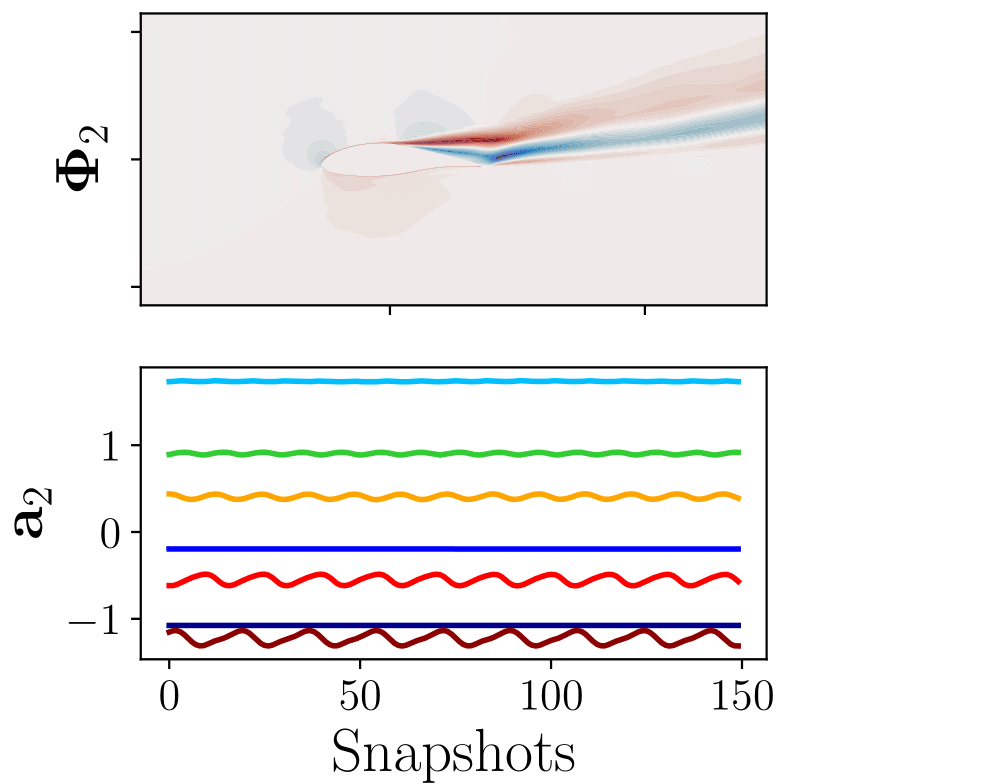} & \includegraphics[trim = 0.1cm 0 0 0, clip,  height=4.6cm]{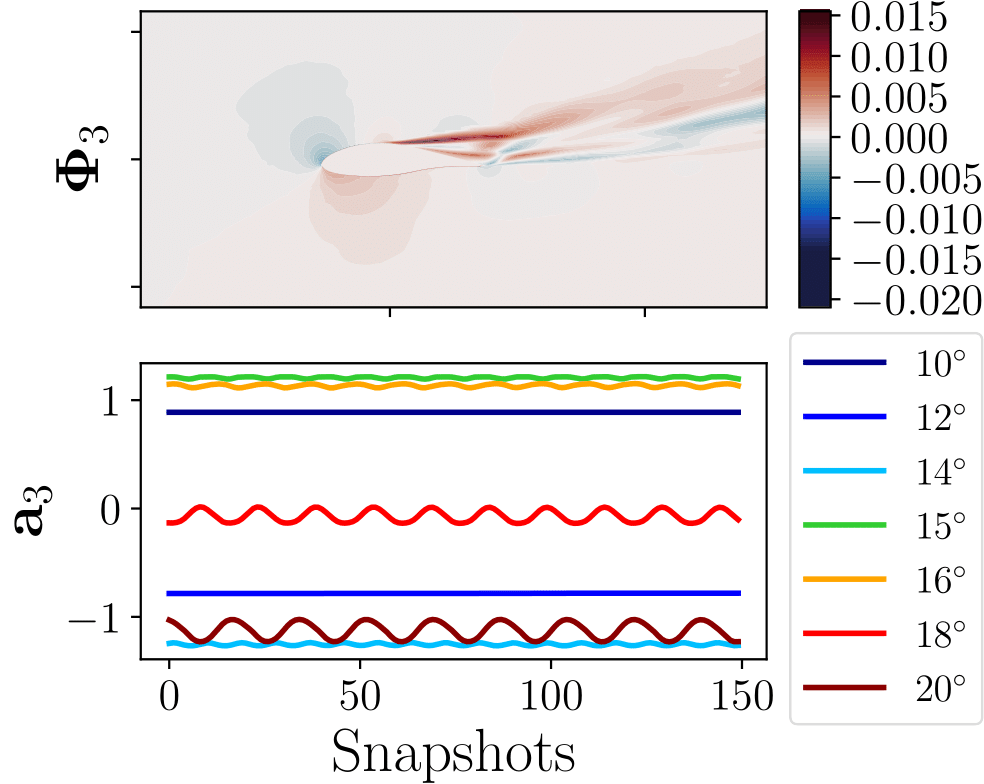} \\
			     (a) Mode 1 ($34.6\%$) & (b) Mode 2 ($10.2\%$) & (c) Mode 3 ($6.2\%$) \\
			     \includegraphics[trim = 0.1cm 0 2.7cm 0, clip,  height=4.6cm]{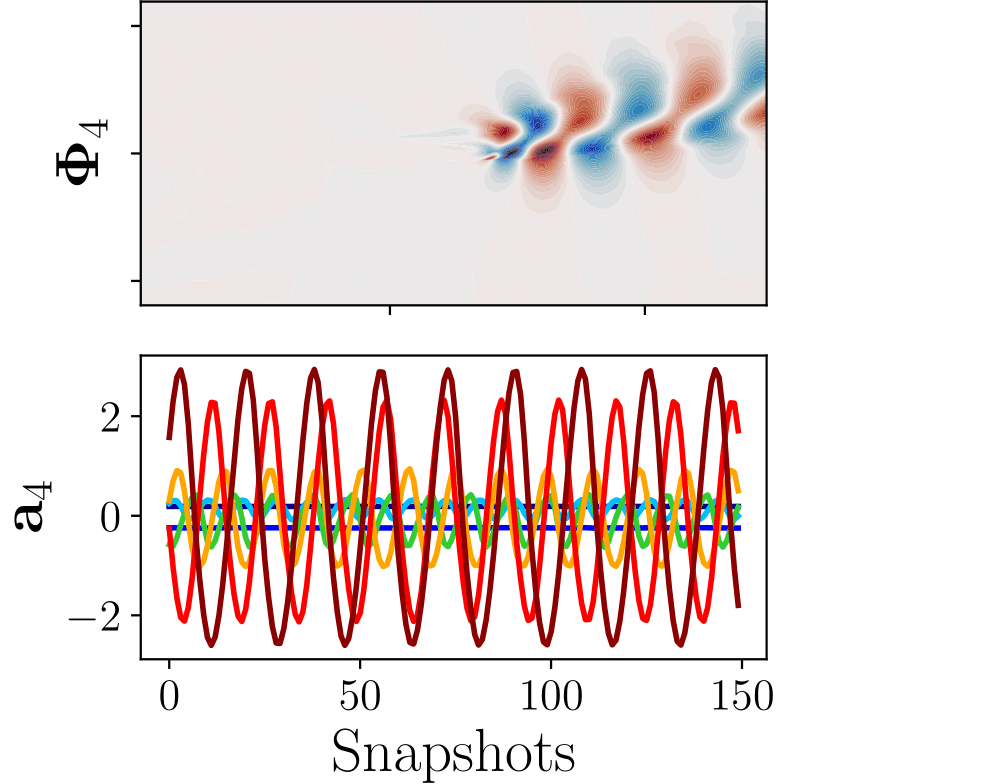} & \includegraphics[trim = 0.1cm 0 2.7cm 0, clip,  height=4.6cm]{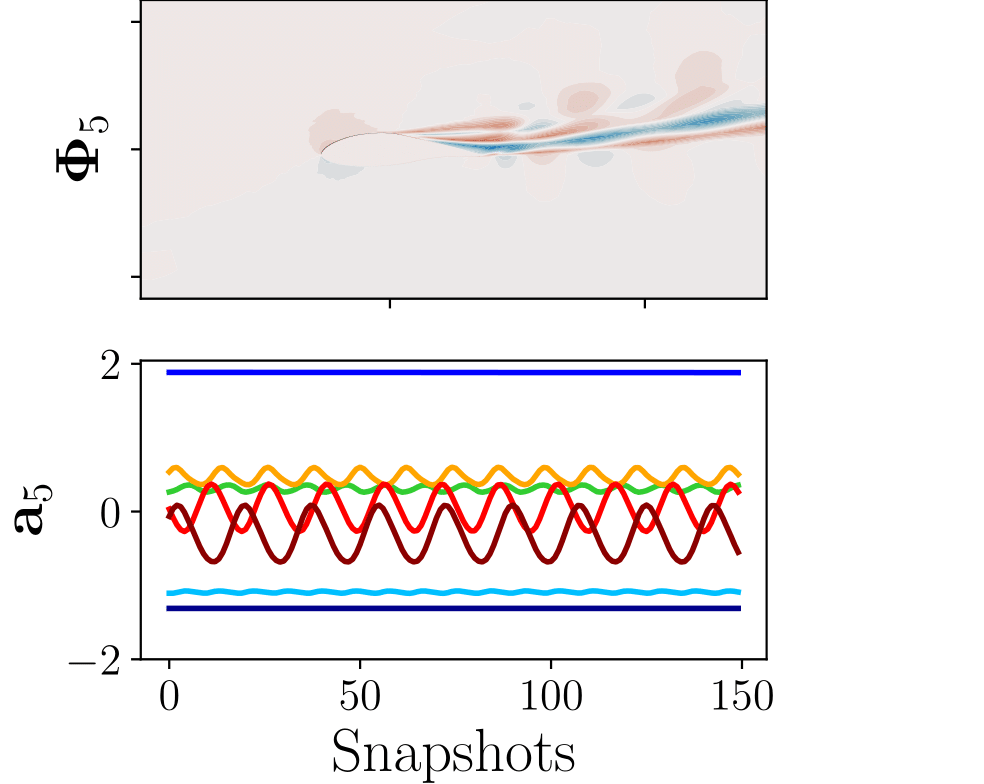} & \includegraphics[trim = 0.1cm 0 0 0, clip,  height=4.6cm]{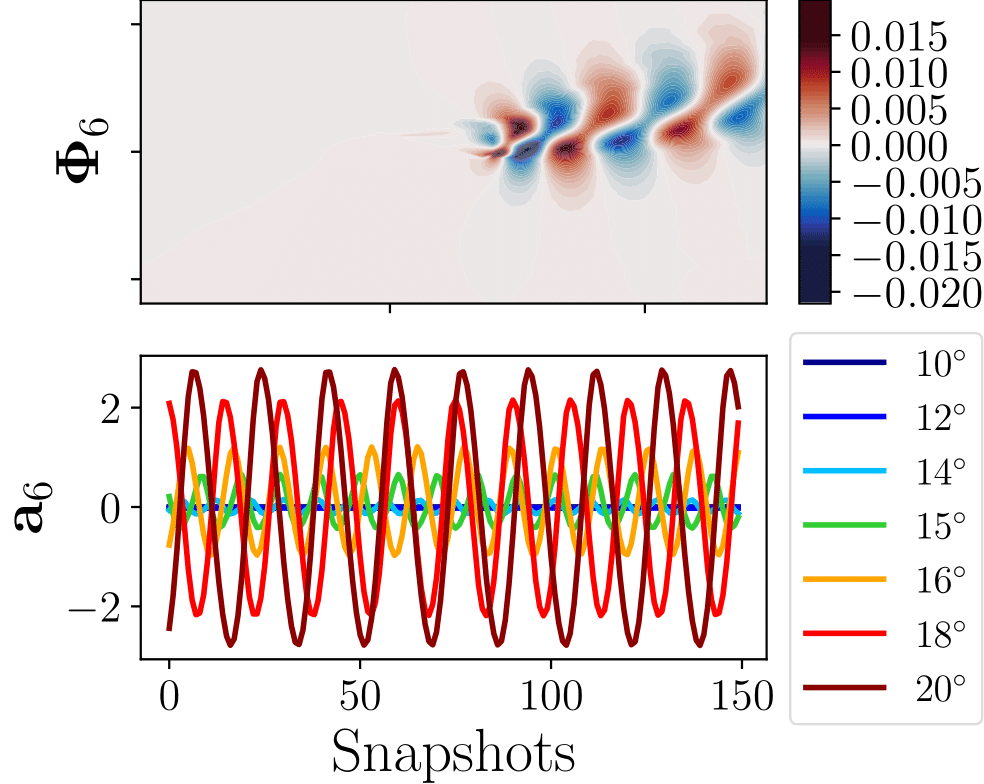} \\
			     (d) Mode 4 ($5.3\%$) & (e) Mode 5 ($4.9\%$) & (f) Mode 6 ($4.4\%$) \\
			
			     \end{tabular}
		     \caption{Sets of POD modes $\mathbf{\Phi_i}$ (top) and their associated POD coefficients $\mathbf{a_i}$ (bottom) for $i=1,...,6$ (a,...,f). Their contribution to the total variance is denoted in parenthesis in captions}
		     \label{fig:U_POD_modes}
		 \end{figure}

	Figure \ref{fig:U_POD_modes} shows the first six most energetic POD spatial modes $\phi_n(\mathbf{x})$, $n=1, \ldots 6$ and their amplitudes $a_n$ which are plotted separately for each angle of attack. 
	The six spatial modes can be split into two groups : on the one hand, modes 1, 2, 3 and 5 are important in the shear layer and separation zone while on the other hand modes 4 and 6 are dominant in the wake.
	The energy level of the latter modes is similar, and their shape is consistent with a pair of vortex shedding modes.
	It can be further observed that the amplitudes of the modes in the first group are characterized by non-zero time-averaged values which are distinct for each angle of attack, which indicates that they correspond to time-independent field deformation due to the angle of attack.
	In contrast the time-average of the amplitudes in the second group is zero and the amplitudes are characterized by large oscillations which increase with separation.
	A criterion can thus be defined to determine whether a POD mode $a_n$ is {\it steady} or {\it unsteady} by comparing its time average for each angle of attack $\overline{a}_n(\text{AoA})$ with the amplitude of its variations. 
	Specifically,  a mode will be considered to be steady (resp. unsteady) if :
	\begin{equation}
		\label{eq:mode_criterion}
		\max\limits_{\text{AoA}} \lvert \max\limits_{N_t} \mathbf{a}_n - \min\limits_{N_t} \mathbf{a}_n \rvert > \max\limits_{\text{AoA}} \lvert \overline{\mathbf{a}_n} \rvert.
	\end{equation}
	This makes it possible to split POD modes into steady and unsteady modes, which will be indexed respectively with the suffixes $st$ and $un$.
	A discussion on the relevance and robustness of this splitting is provided in Sec. \ref{sec:modal_splitting_robustness}.

	\subsubsection{Construction of the model}
	\label{sec:SNN_architecture}
	
	Given an input vector of pressure measurements of size $p$, the model learns the projection onto the POD basis by estimating the $r$ most energetic POD coefficients $\hat{a}_{i}$, which are then combined to reconstruct the flow field of size $N_s$ using :
	\begin{equation}
		\label{eq:POD_recons}
		\hat{u}(\mathbf{x}) = \langle u(\mathbf{x}) \rangle + \sum_{1\le n \le r} \lambda_n^{1/2} \hat{a}_n  \phi_n (\mathbf{x}) 
	\end{equation}
	where $ \langle u(\mathbf{x}) \rangle$, $\lambda_n$ and  $\phi_n$ are determined from the training set.
	A preliminary analysis conducted on a single AoA configuration (not shown) indicated that increasing the number of POD modes generally enhances the reconstruction accuracy; however, the improvement becomes marginal for $r \geq 50$ retained modes. For this reason, the value of  $r=50$ POD modes, corresponding to a compression ratio of $96.75\%$, was chosen for the present study. 
	
	The amplitudes to be learnt $\hat{a}_n$ are split into two output vectors consisting of steady and unsteady modes—according to previous section—which are independently learnt by two networks based on the shallow decoder architecture outlined in \cite{erichson_shallow_2020}, which we first briefly review. 
	A Shallow Neural Network (SNN) is a neural network consisting  of $L$ fully-connected hidden layers (with $L \le 3$ in general). 
	Given an input  $\mathbf{s}$, the output of the model is defined as 
	$$
	\mathcal{F}(\mathbf{s}; \mathbf{w}) = \mathbf{w^L} g (\mathbf{w^{L-1}}...  g(\mathbf{w^{1} s})),
	$$
	where $g$ is a non-linear activation function and $\mathbf{w}$ the set of weights matrices.
	Here $L=3$ and $g$ is the ReLu function. 
	The model used in the present study relies on two SNNs with identical  architectures. 
	The robustness of each network is improved by the addition of a dropout with a drop ratio of $10\%$. 
	An Adaptative Moment Estimation (ADAM) optimizer is used along with a regularization method (weight decay) to prevent overfitting. 
	Training batches are shuffled and mini-batch size is set to $128$. 
	Both networks are trained over 4000 epochs but training may be shortened by early stopping that acts as a regularizer to avoid overfitting \citep{goodfellow_2016}.
	In practice early stopping is applied  if the error on the testing dataset does not reduce over 100 epochs. 
	The choice of the optimal set of hyperparameters (i.e. parameters external to the learning process which are directly learnt within estimators) is determined by using Optuna hyperparameter optimization software framework based on Bayesian optimization \citep{optuna_2019}. The list of optimized hyperparameters and their associated values is given in Table \ref{tab:hyperparameters}.
	
	\begin{center}
		\begin{table}[h]
			\centering
			\begin{tabular}{ccccc}
				\hline
				\hline
				Learning rate & Learning rate change & Weight decay & Weight decay change & Dropout probability\\
				\hline
				\num{1.5e-2} & \num{7e-5} & 0.8 & 0.7 & 0.1\\
			\end{tabular}\hspace{0pt}%
			\begin{tabular}{cccc}
				\hline
				\hline
				Input layer size & Hidden layer size 1 & Hidden layer size 2 & Output layer size \\
				\hline
				$p$ & $r$ & $1.4 r$ & $r$ \\
			\end{tabular}\hspace{0pt}%
			\caption{List of the tuned hyperparameters and their optimal values}
			\label{tab:hyperparameters}
		\end{table}
	\end{center}
	
	Each SNN is trained to minimize the $L_2$-norm of the error between the reduced state and its estimation, weighted by the 
	associated eigenvalue (contribution of the mode to the variance), so that 
	the loss is given by :
	\begin{equation}
		\begin{gathered}
			\label{eq:loss}
			\mathcal{L}_{POD} = \left\Vert \mathbf{S} \left[ \mathbf{A}^k - \mathcal{F}(\mathbf{s}^k; \mathbf{w}) \right] \right\Vert_2 = 
			\sum_{1 \le n \le r} \lambda_n \sum_{1 \le m \le N_t N_{\text{AoA}}} \| a_n^m- \hat{a}_n^m \|^2
		\end{gathered}
	\end{equation}
	
	The singular values $\mathbf{S}$ included in the loss function are not learned by the model, but are directly extracted from the POD decomposition of the training dataset. In a sense, this formulation can be interpreted as a physics-based loss where the contribution of each coefficient to the loss represents its contribution to the total variance. 

	The test set consists of unseen snapshots  for  angles of attack  both within and outside $I_{train}$. 
	Specifically, three new angles  were considered: $15.5^\circ, 17^\circ, 19^\circ$ so the total number of angles in the test set is $N_{\text{AoA}}^{test}=10$. 
	For each AoA $\in I_{test}$ a sequence of $N_{t}^{test}=50$ snapshots was selected to constitute the test set presented in the paper. 
	Results did not change significantly for larger numbers or random selections of snapshots. 
	
	To evaluate the model accuracy on the test set, the following metrics  were used:
	\begin{itemize}
		\item the normalized mean-square error (MSE) $E_{test}$, given by :
		\begin{equation}
			E_{test} = \frac{1}{N_{\text{AoA}}^{test}N_{t}^{test}} \sum_{\substack{1 \le i \le N_{\text{AoA}}^{test} \\ 1 \le k \le N_t^{test}}} \frac{\left\Vert \mathbf{u}^{k,i} - \widehat{\mathbf{u}}^{k,i} \right\Vert_2}{\left\Vert \mathbf{u}^{k,i} \right\Vert_2}
		\end{equation}
		
		\item the fluctuation error  $\varepsilon_{test}$, based on the fluctuating part of the velocity field $u'$ (with respect to the average taken over
		the training set), defined as
		\begin{equation}
			\varepsilon_{test} = \frac{1}{N_{\text{AoA}}^{test}N_{t}^{test}} \sum_{\substack{1 \le i \le N_{\text{AoA}}^{test} \\ 1 \le k \le N_t^{test}}} \frac{\left\Vert \mathbf{u'}^{k,i} - \widehat{\mathbf{u'}}^{k,i} \right\Vert_2}{\left\Vert \mathbf{u'}^{k,i} \right\Vert_2}
		\end{equation}
		
		By construction, the fluctuation error is larger than the MSE. It is also independent from the fluctuation to mean ratio, which can bias the MSE. 
		
		\item the projection error $\varepsilon_{test}^{\mathbf{\Phi}}$  measures the relative difference between the reconstruction  
		and the projection of the ground truth velocity field onto the POD basis of $r$ modes:    
		\begin{equation}
			\varepsilon_{test}^{\mathbf{\Phi}} = \frac{1}{N_{\text{AoA}}^{test}N_{t}^{test}} \sum_{\substack{1 \le i \le N_{\text{AoA}}^{test} \\ 1 \le k \le N_t^{test}}} \frac{\left\Vert \mathbf{u'}_{\mathbf{\Phi}}^{k,i} - \widehat{\mathbf{u'}}^{k,i} \right\Vert_2}{\left\Vert \mathbf{u'}_{\mathbf{\Phi}}^{k,i} \right\Vert_2}.
		\end{equation}
		where $ \mathbf{u'}_{\mathbf{\Phi}}^{k,i}= \sum_{1 \le n \le r} a_n^{k,i} \Phi_n $ is the projection of the ground truth field onto the POD basis of
		the training set, which represents the best approximation achievable by the SNN-POD. 
		Comparing the fluctuation and the projection error therefore provides a measure of  the suitability of the POD training basis. 
		
	\end{itemize}
	
	\section{Results}
	\label{sec:results}
	
	In this section,  the performance of the SNN-POD model trained on the full range of angles of attack is evaluated on the test set. 
	The robustness of the model is then examined through a sensitivity analysis (Sec. \ref{sec:sensor_location_sensitivity}) of the sensor placement strategy described in Sec. \ref{sec:sensors_placement_description}, followed by a study of the influence of the modal splitting approach (Sec. \ref{sec:modal_splitting_robustness}) on the model performance. 
	Finally, Sec. \ref{sec:specific_model} describes how combining SNN-POD models trained on different subsets of angles of attack can locally improve the reconstruction.
	
	\subsection{Evaluation of SNN-POD}
	\label{sec:evalsnnpod}
	
\begin{figure}[h]
		\centering
		\includegraphics[width=\linewidth]{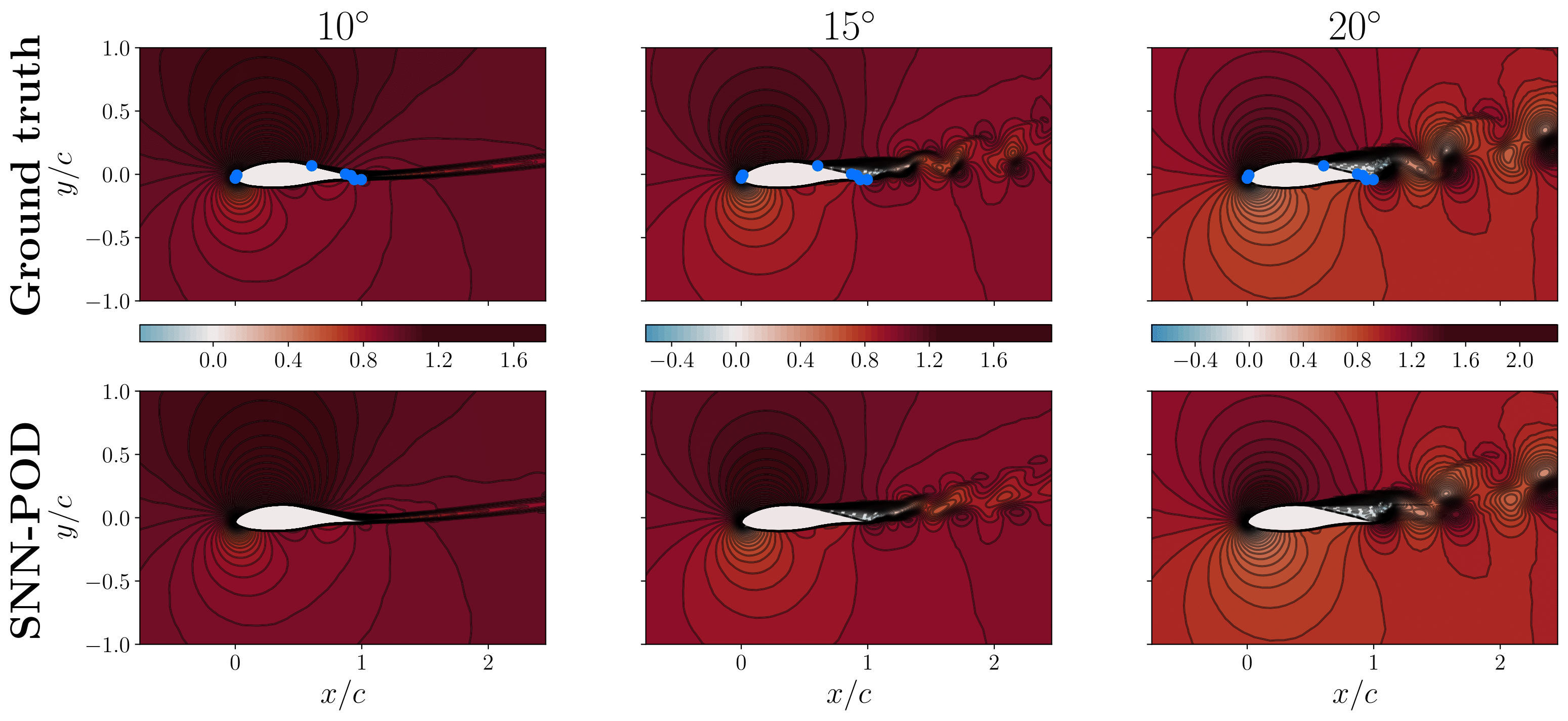}
		\caption{Comparison of streamwise velocity $u(\mathbf{x},t_s)/U_\infty$ contours at snapshot $t_s=150$s of the ground truth obtained from URANS (left) and its SNN-POD estimation (right) for three AoA : $10^\circ$ (left), $15^\circ$ (center) and $20^\circ$ (right). The locations of the $p=7$ pressure sensors on the blade are indicated with blue markers}
		\label{fig:SNN_POD_evaluation_seen_AoA_U}
	\end{figure}

\begin{figure}[h!]
			\centering
		\begin{subfigure}{0.32\textwidth}
			\centering
			\includegraphics[width=\linewidth]{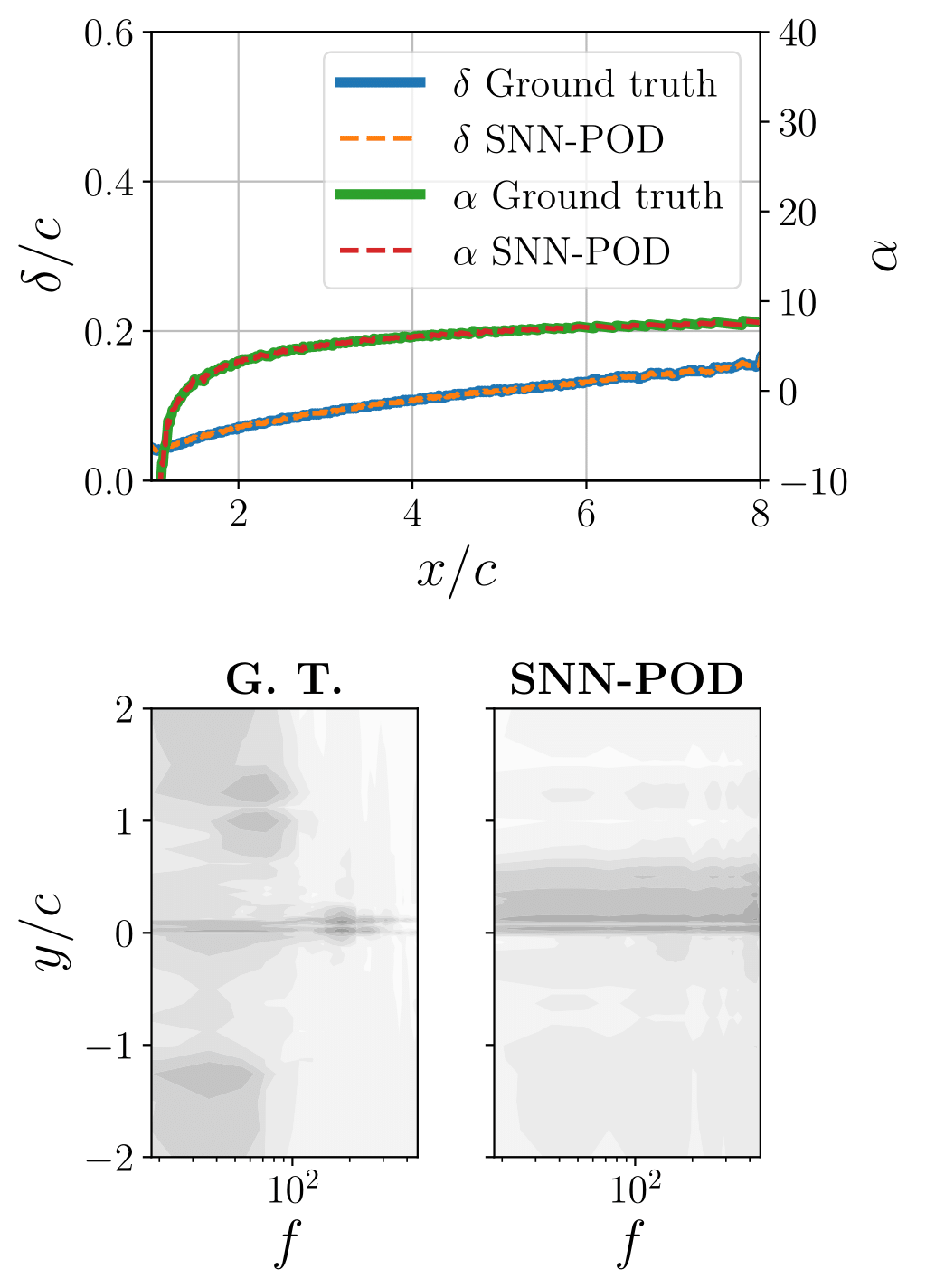}
			\caption{AoA = $10^\circ$}
		\end{subfigure}
		\begin{subfigure}{0.32\textwidth}
			\centering
			\includegraphics[width=\linewidth]{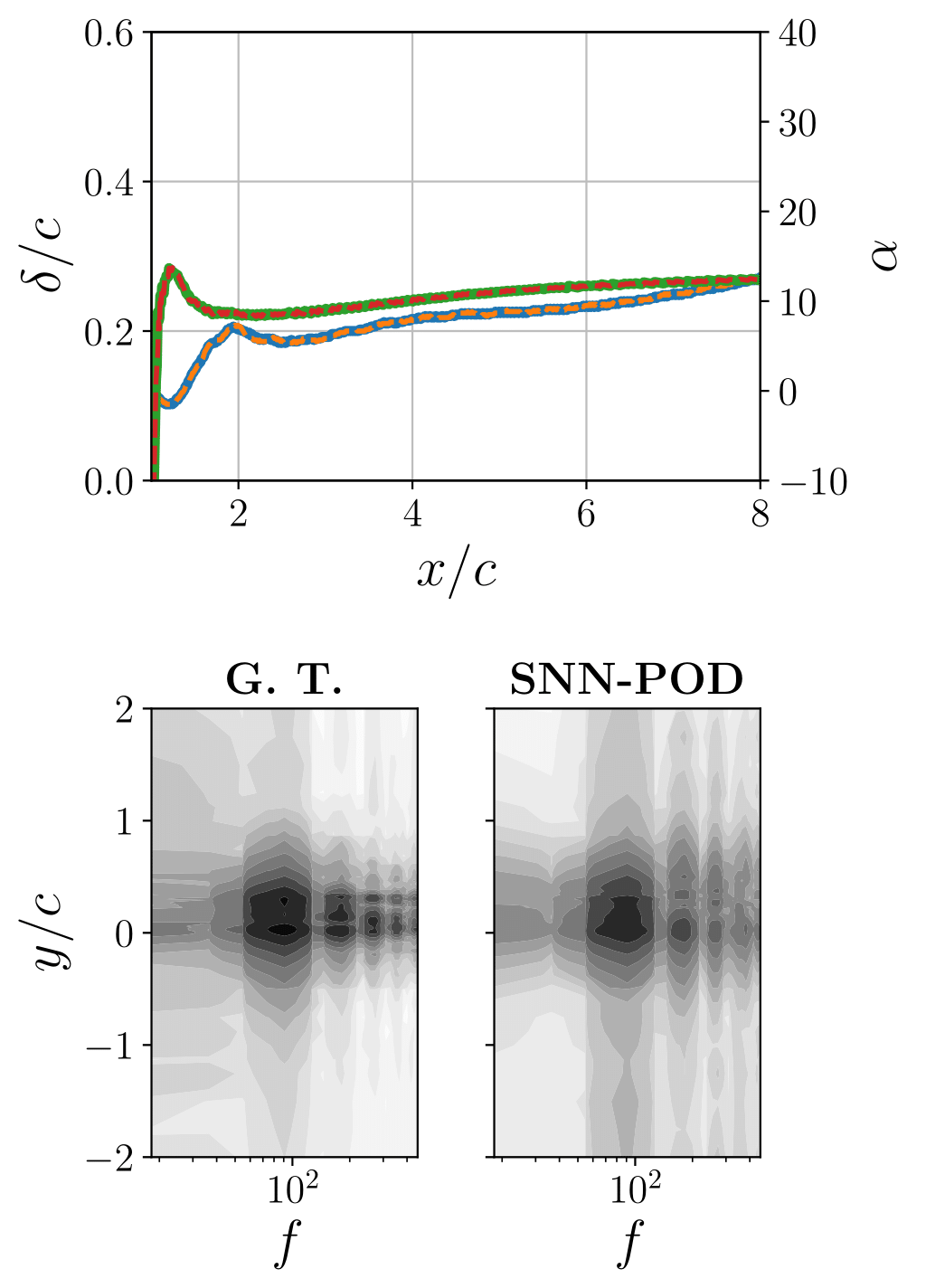}
			\caption{AoA = $15^\circ$}
		\end{subfigure}
		\begin{subfigure}{0.32\textwidth}
			\centering
			\includegraphics[width=\linewidth]{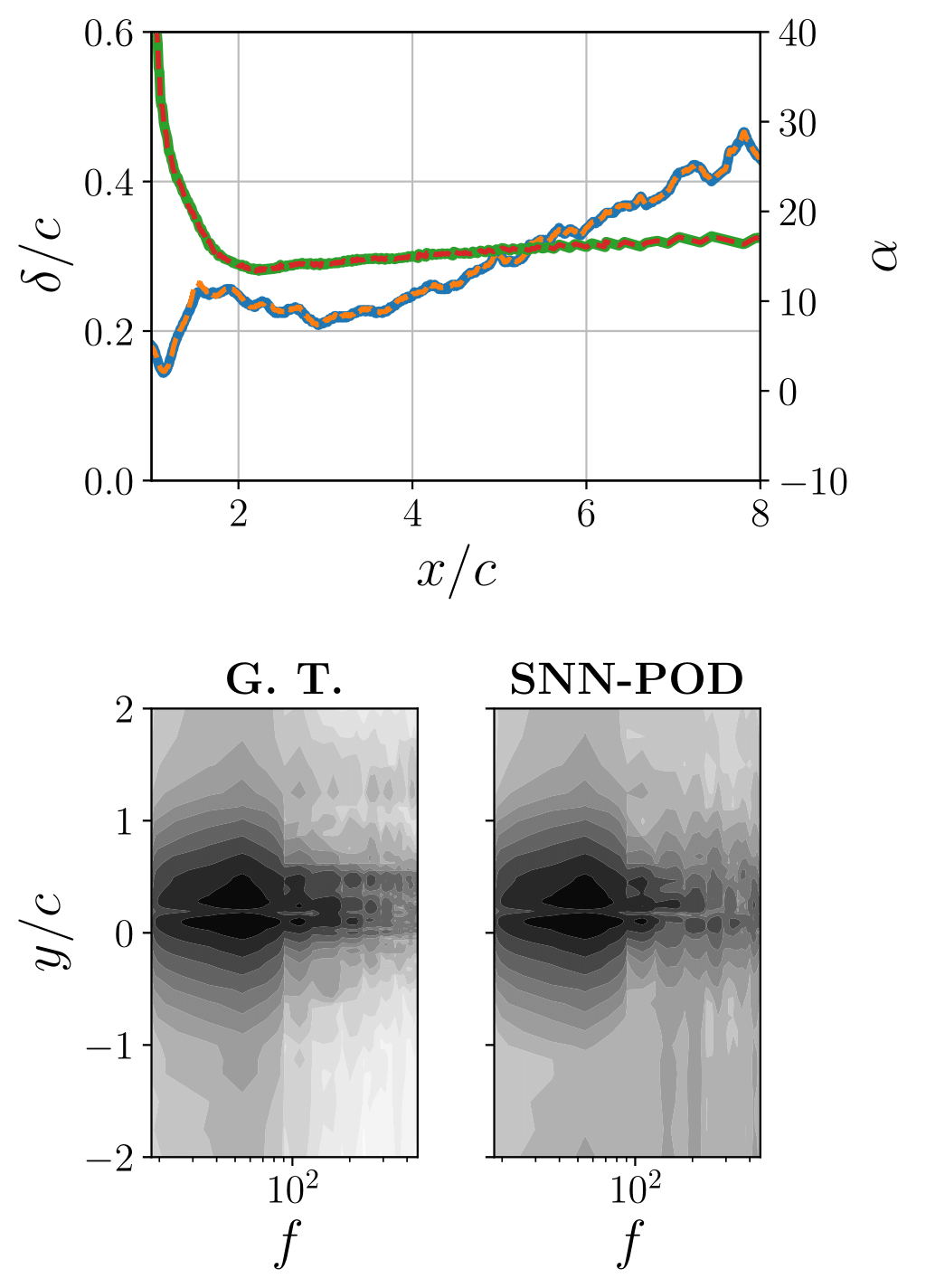}
			\caption{AoA = $20^\circ$}
		\end{subfigure}
		\caption{Top plots denote the wake widths and wake deflection angles comparison between the ground truth and the estimation for three AoA : $10^\circ$ (left), $15^\circ$ (center) and $20^\circ$ (right). Bottom plots correspond to the spectrograms of the pre-multiplied velocity spectra $f S_u$ of the ground truth  (G.T., left) and estimation (SNN-POD, right) at the $x/c = 2$ section. The same logarithmic colormap is applied for both spectrograms}
		\label{fig:SNN_POD_evaluation_seen_AoA_wake}
	\end{figure}

\begin{figure}[h]
		\centering
		\includegraphics[width=\linewidth]{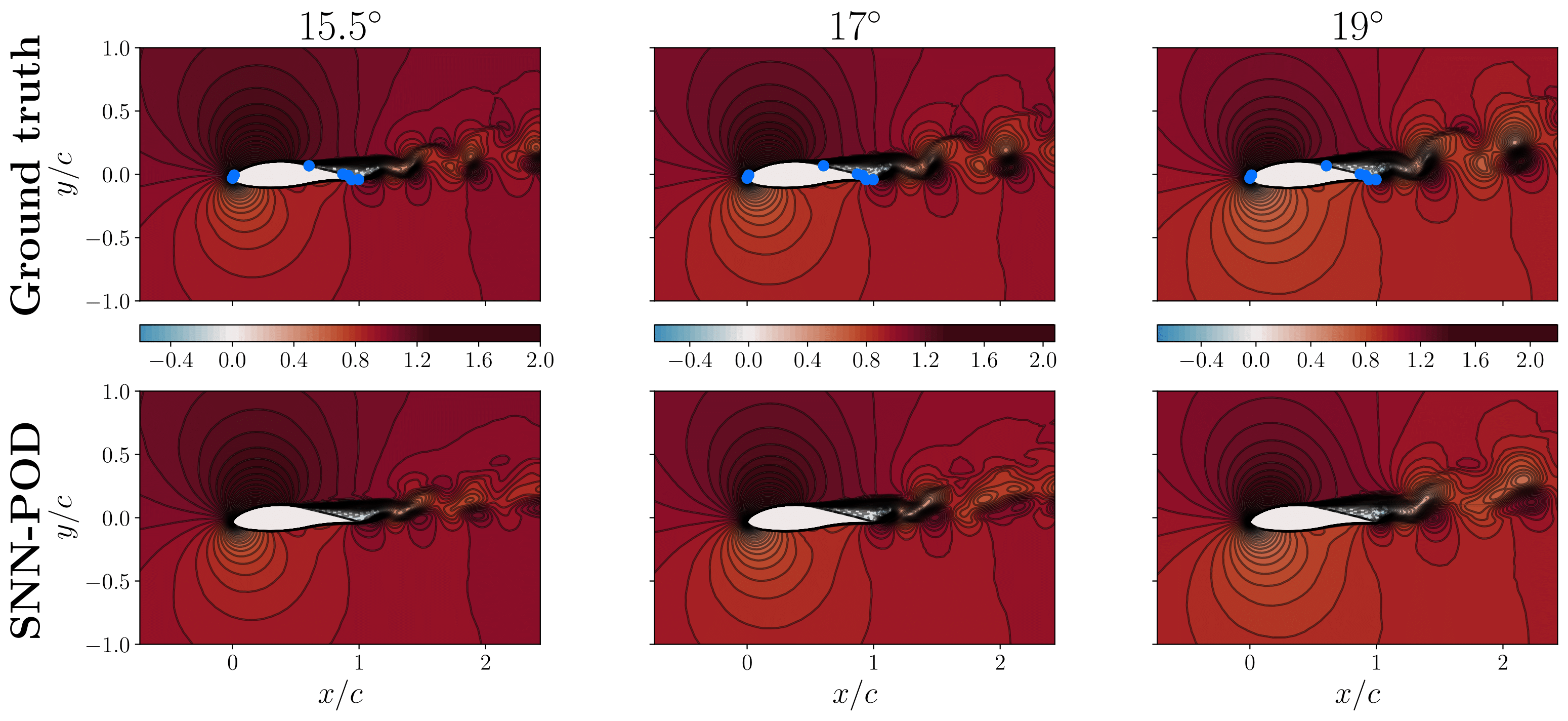}
		\caption{Comparison of streamwise velocity $u(\mathbf{x},t_s)$ contours at snapshot $t_s=150$s of the ground truth obtained from URANS (first plot from left to right) and its SNN-POD estimation (second plot) for three interpolated AoA : $15.5^\circ$ (left), $17^\circ$ (center) and $19^\circ$ (right). The locations of the $p=7$ pressure sensors on the blade are indicated with blue markers}
		\label{fig:SNN_POD_evaluation_unseen_AoA_U}
	\end{figure}

\begin{figure}[h!]
		\centering
		\begin{subfigure}{0.32\textwidth}
			\centering
			\includegraphics[width=\linewidth]{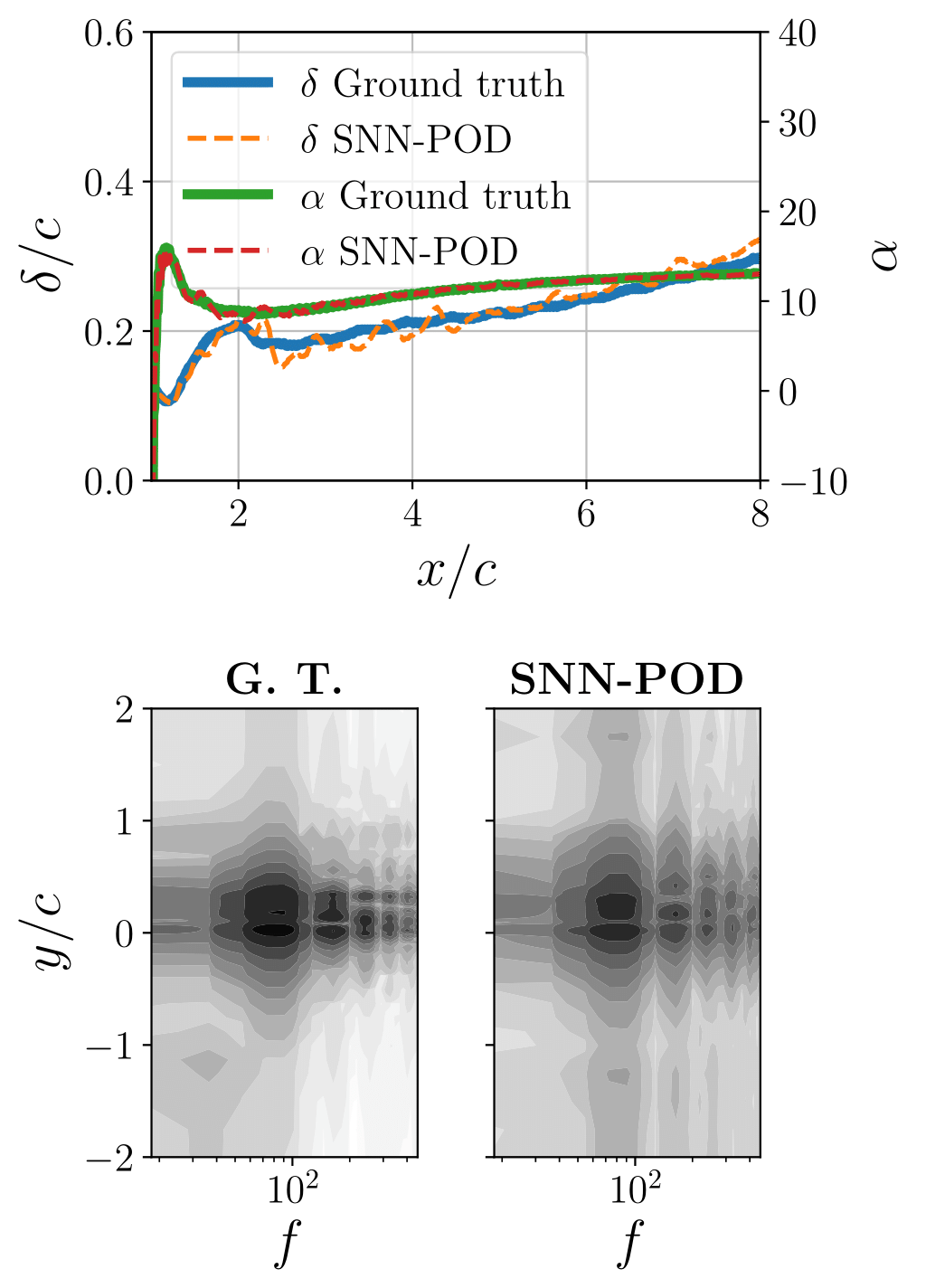}
			\caption{AoA = $15.5^\circ$}
		\end{subfigure}
		\begin{subfigure}{0.32\textwidth}
			\centering
			\includegraphics[width=\linewidth]{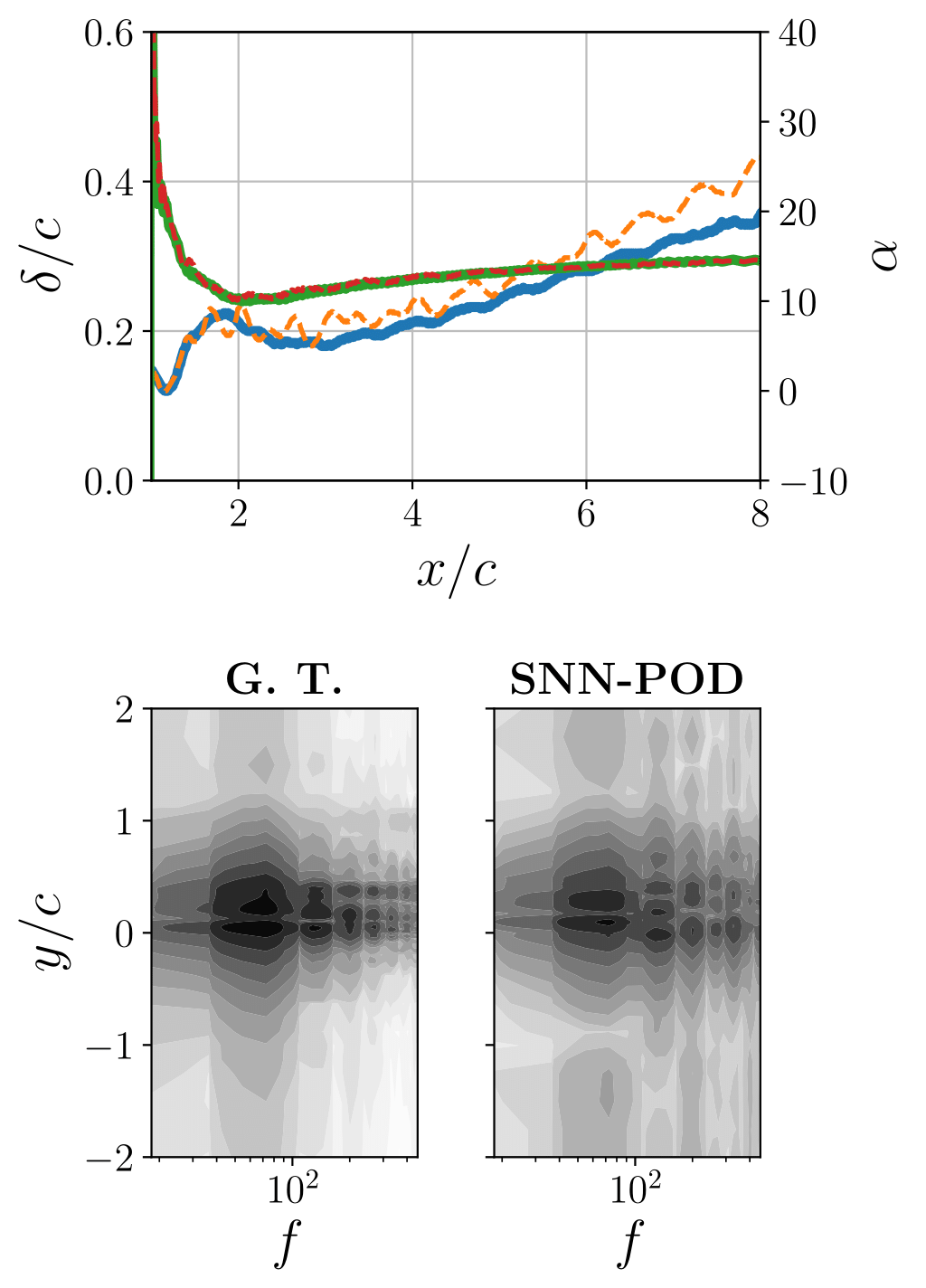}
			\caption{AoA = $17^\circ$}
		\end{subfigure}
		\begin{subfigure}{0.32\textwidth}
			\centering
			\includegraphics[width=\linewidth]{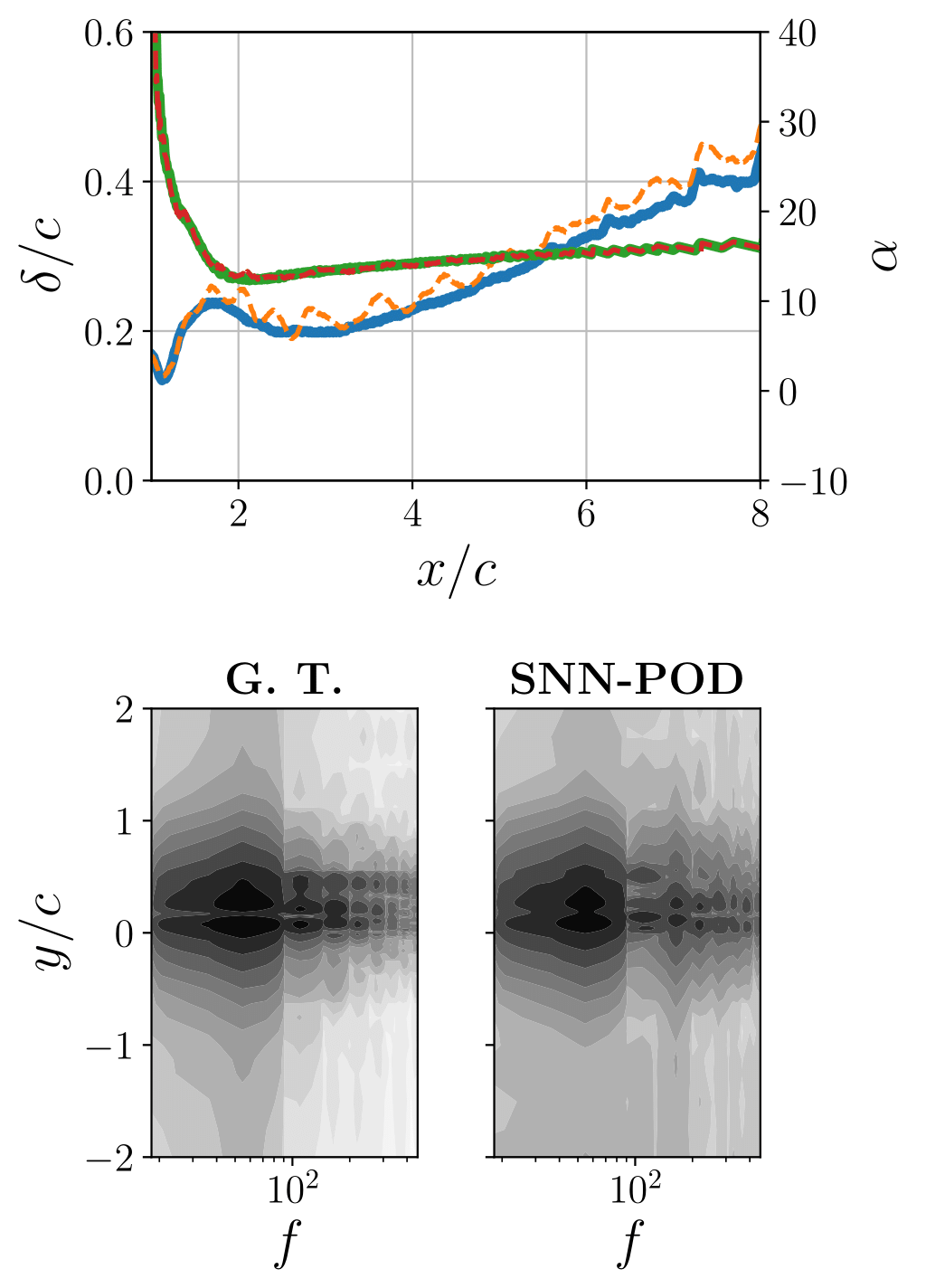}
			\caption{AoA = $19^\circ$}
		\end{subfigure}
		\caption{Wake widths and wake deflection angles comparison between the ground truth and the estimation (top) for three unseen AoA : $15.5^\circ$ (left), $17^\circ$ (center) and $19^\circ$ (right). Bottom plots correspond to the spectrograms of the pre-multiplied velocity spectra $f S_u$ of the ground truth  (G.T., left) and estimation (SNN-POD, right) at the $x/c = 2$ section. The same logarithmic colormap is applied for both spectrograms}
		\label{fig:SNN_POD_evaluation_unseen_AoA_wake}
	\end{figure}

	This section evaluates the ability of the SNN-POD model to estimate both the POD coefficients and the instantaneous velocity field $u$ from the URANS simulations, denoted thereafter as the “ground truth”.
	As mentioned earlier, the test set includes angles sampled during training as well as angles which are not part of the original discretization (interpolated angles). 
	Figs. \ref{fig:SNN_POD_evaluation_seen_AoA_U}-\ref{fig:SNN_POD_evaluation_unseen_AoA_wake} compare characteristic features of the true and reconstructed flow fields for three sampled $(10^\circ, 15^\circ, 20^\circ)$ and three interpolated angles $(15.5^\circ, 17^\circ, 19^\circ)$.
	Each column corresponds to an angle of attack.
	Figs. \ref{fig:SNN_POD_evaluation_seen_AoA_U} and \ref{fig:SNN_POD_evaluation_unseen_AoA_U} provide a comparison of a ground truth snapshot (top row) with its reconstruction (bottom row) for sampled and interpolated angles, respectively. 
	In all cases a very good agreement is observed for the main features of the flow, although some slight differences are noticeable for the interpolated angles, particularly in the far wake.
	
	A more quantitative assessment can be made by examining the top rows of Figs. \ref{fig:SNN_POD_evaluation_seen_AoA_wake} and \ref{fig:SNN_POD_evaluation_unseen_AoA_wake}, which show the average wake width and deflection angles. 
	The wake width is well recovered for the sampled angles (fig. \ref{fig:SNN_POD_evaluation_seen_AoA_wake}). 
	For the interpolated angles (fig. \ref{fig:SNN_POD_evaluation_unseen_AoA_wake}), the wake width is characterized by small oscillations for $x/c>1.5$ and tends to be overpredicted for the two higher angles, although the growth rate $d \delta/dx$ generally appears to be correct.
	In contrast, the deflection angle is well recovered for both sampled and interpolated angles, which shows that the model is able to identify the angle of attack of the configuration from a single set of pressure measurements. This result is particularly important in the context of wind energy, where the estimation of instantaneous wind loads is directly linked to the determination of the AoA.
	
	The bottom lines of Figs. \ref{fig:SNN_POD_evaluation_seen_AoA_wake} and \ref{fig:SNN_POD_evaluation_unseen_AoA_wake} compare the time spectra of the true and estimated velocity fields for sampled and interpolated angles, respectively.
	For all AoAs, the energy levels are accurately predicted, and the characteristic frequencies associated with flow separation are also well captured. 
	The largest discrepancies in the spectrograms are observed for the attached flow case ($10^\circ$), where wake turbulence is low in energy.
	The model is able to predict the frequency shift observed for both sampled and interpolated angles and reproduces the complexity of the spectra, although unsurprisingly the agreement is closer for sampled angles.
	
\begin{figure}[h!]
		\begin{subfigure}{.49\textwidth}
			\centering
			\includegraphics[width=\linewidth]{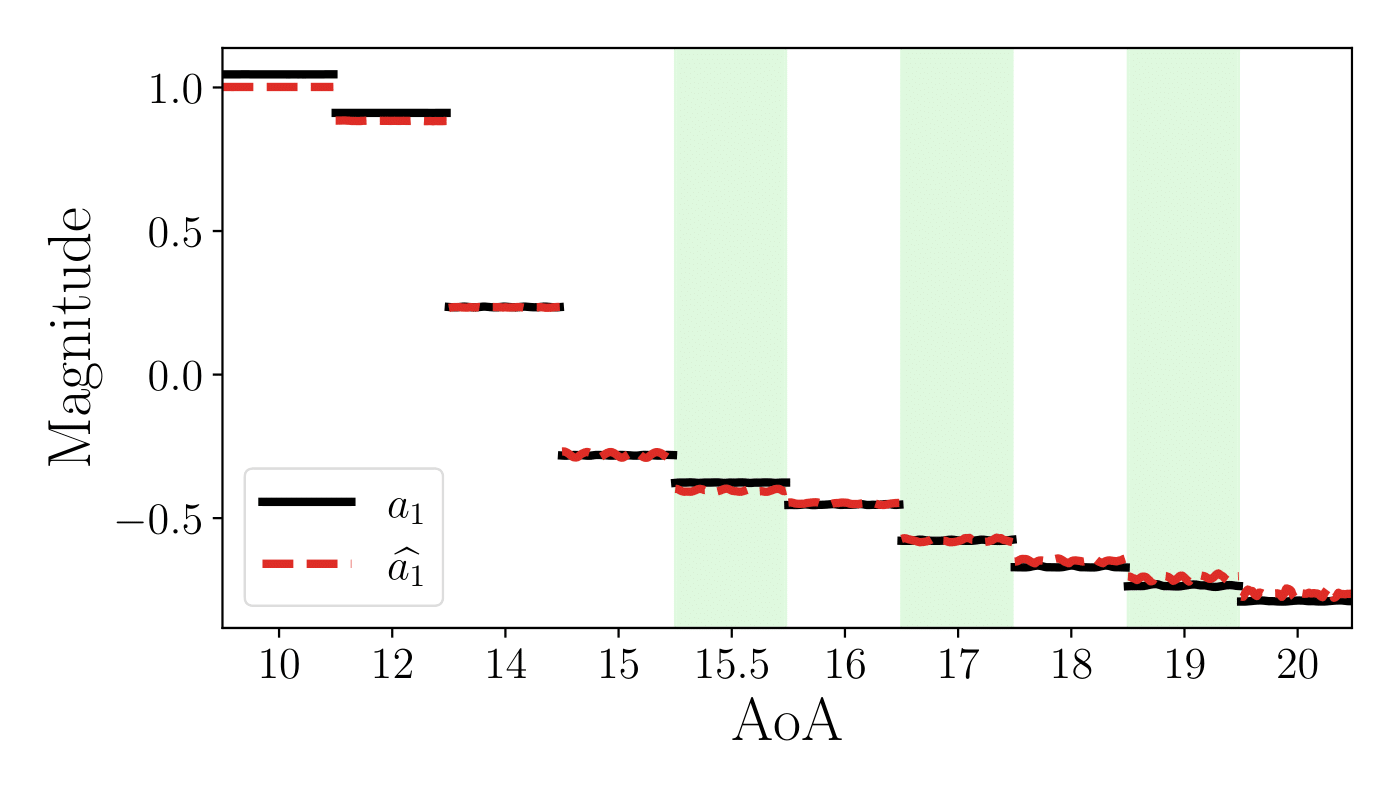}
			\caption{Mode 1}
		\end{subfigure}%
		\begin{subfigure}{.49\textwidth}
			\centering
			\includegraphics[width=\linewidth]{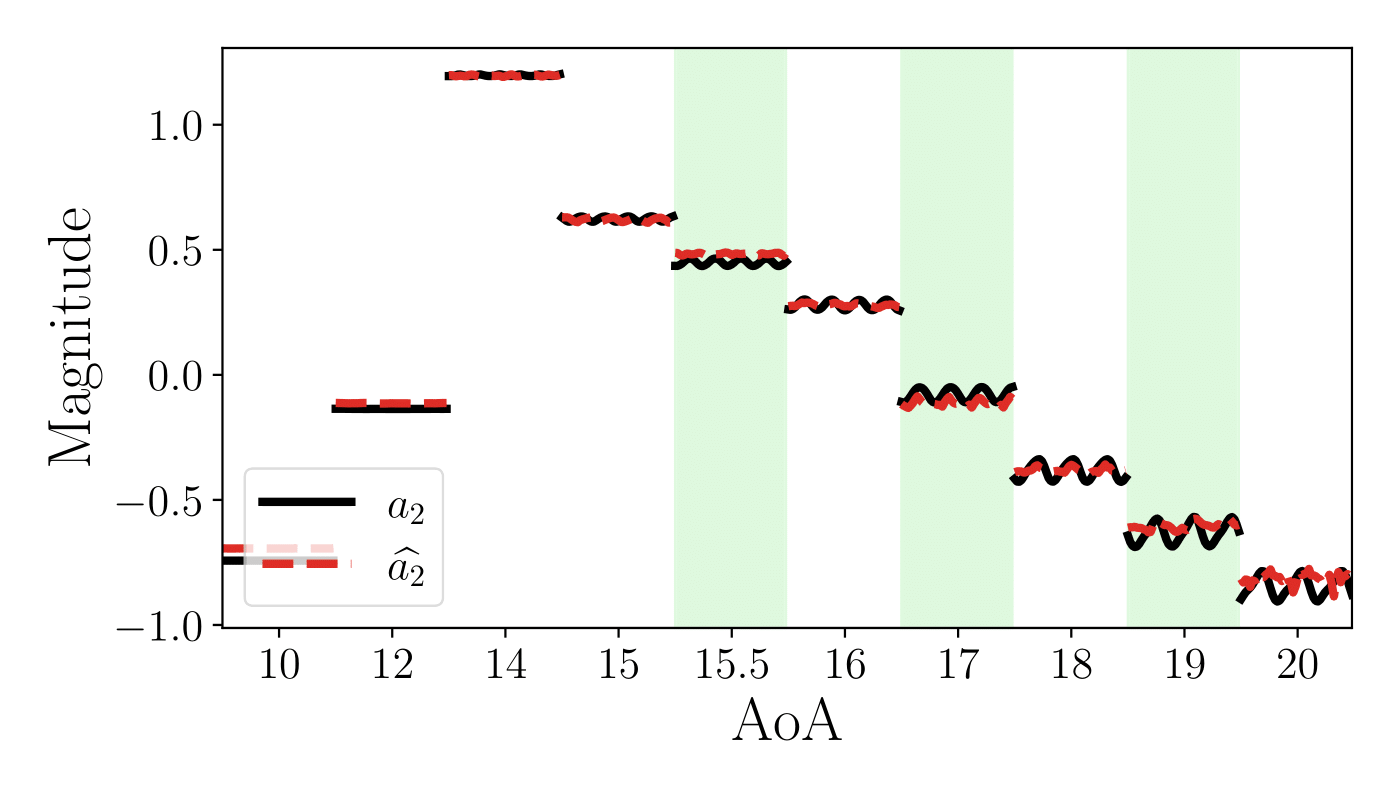}
			\caption{Mode 2}
		\end{subfigure}
		\\
		\begin{subfigure}{.49\textwidth}
			\centering
			\includegraphics[width=\linewidth]{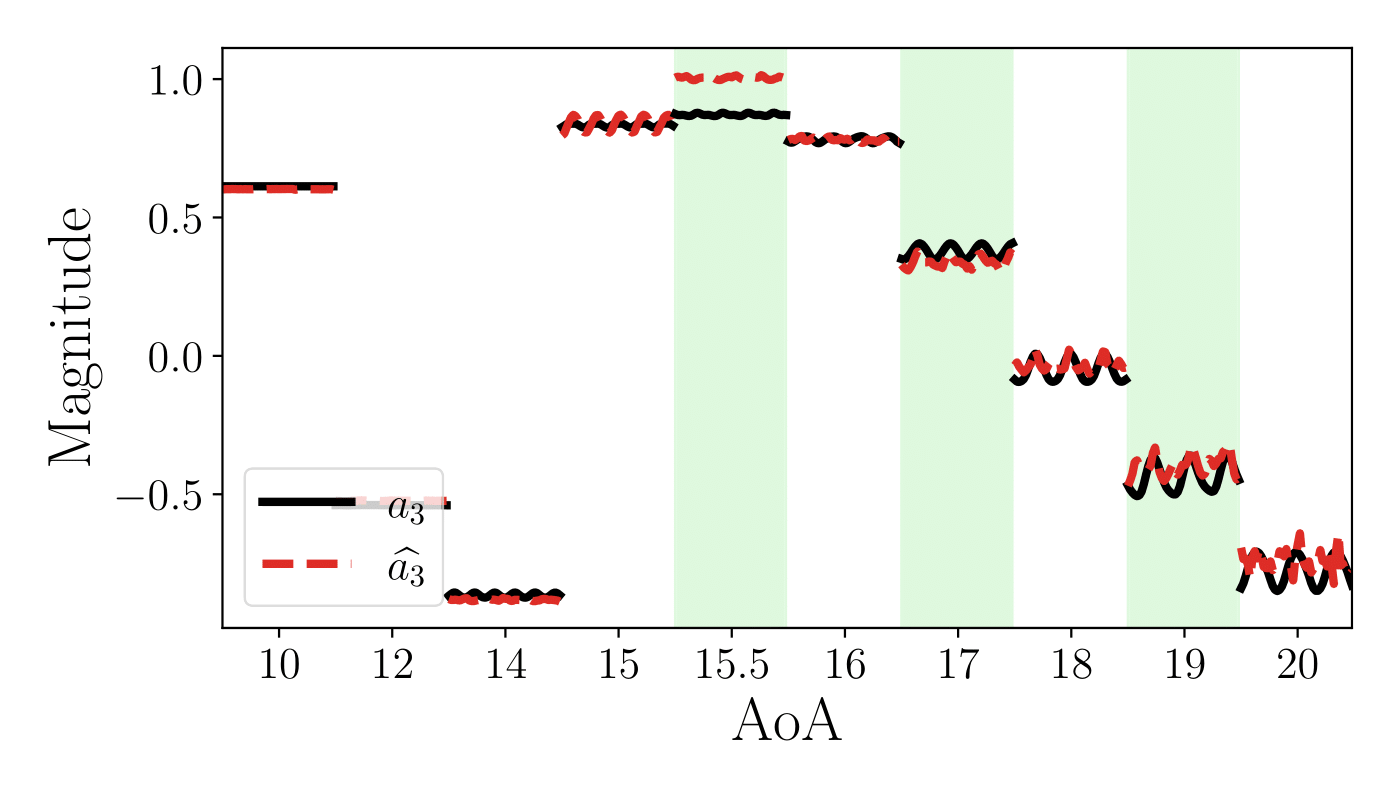}
			\caption{Mode 3}
		\end{subfigure}
		\begin{subfigure}{.49\textwidth}
			\centering
			\includegraphics[width=\linewidth]{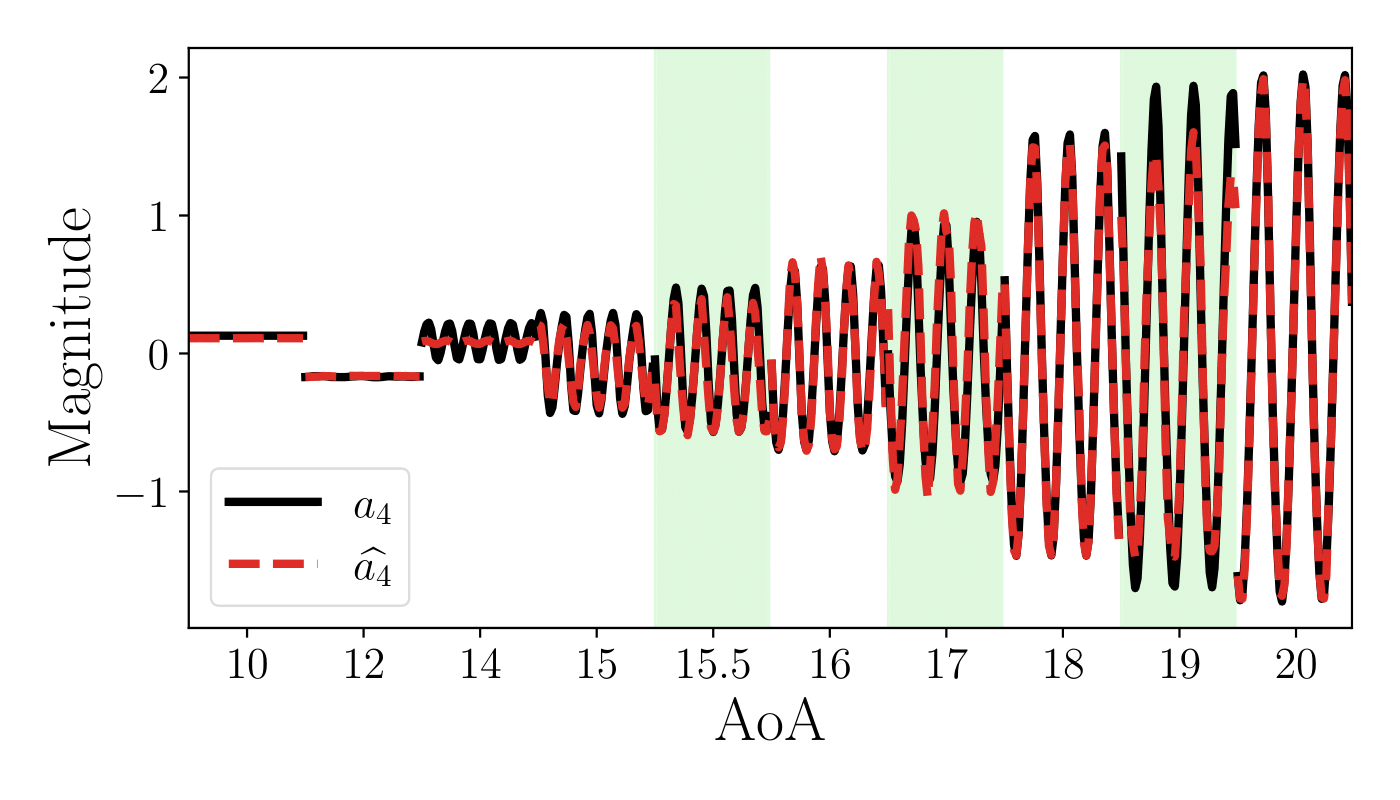}
			\caption{Mode 4}
		\end{subfigure}%
		\\
		\begin{subfigure}{.49\textwidth}
			\centering
			\includegraphics[width=\linewidth]{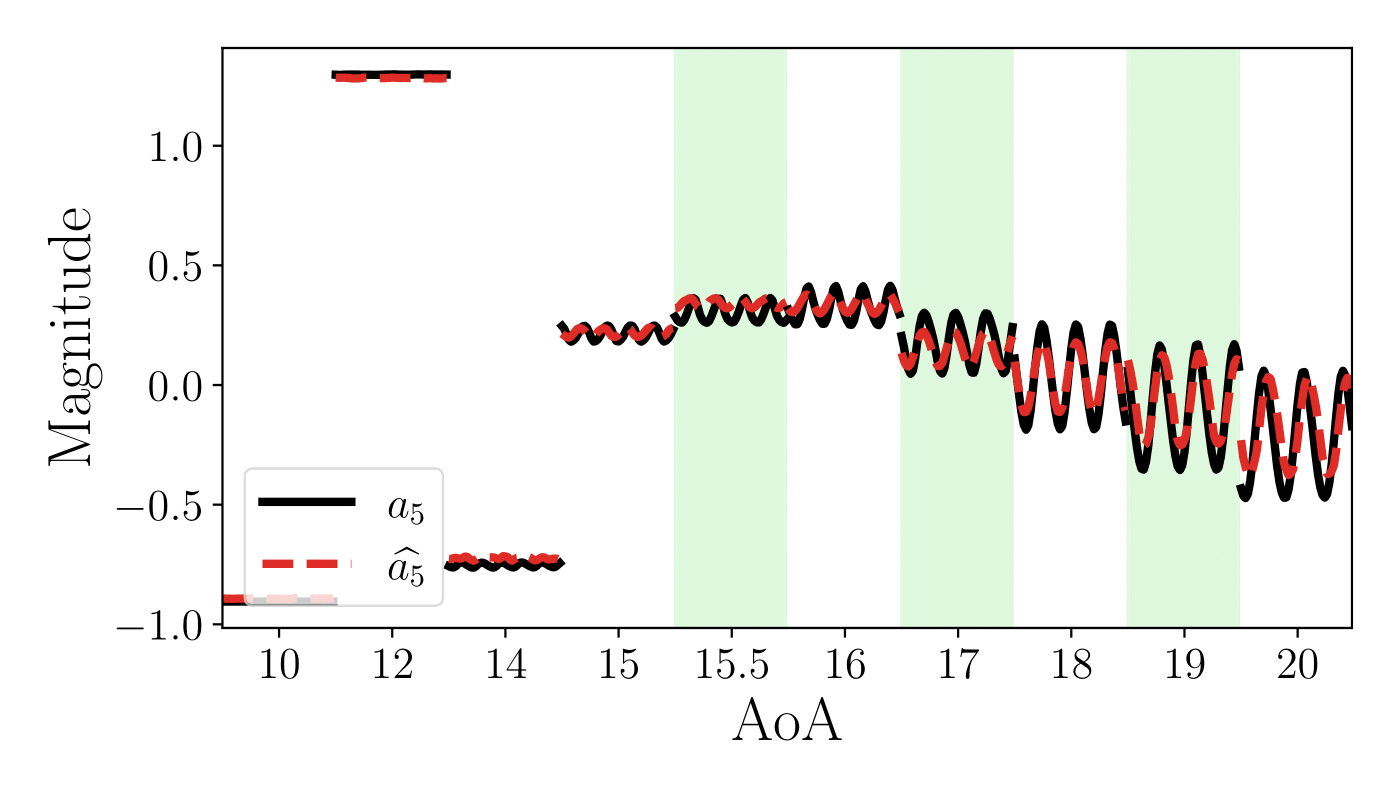}
			\caption{Mode 5}
		\end{subfigure}
		\begin{subfigure}{.49\textwidth}
			\centering
			\includegraphics[width=\linewidth]{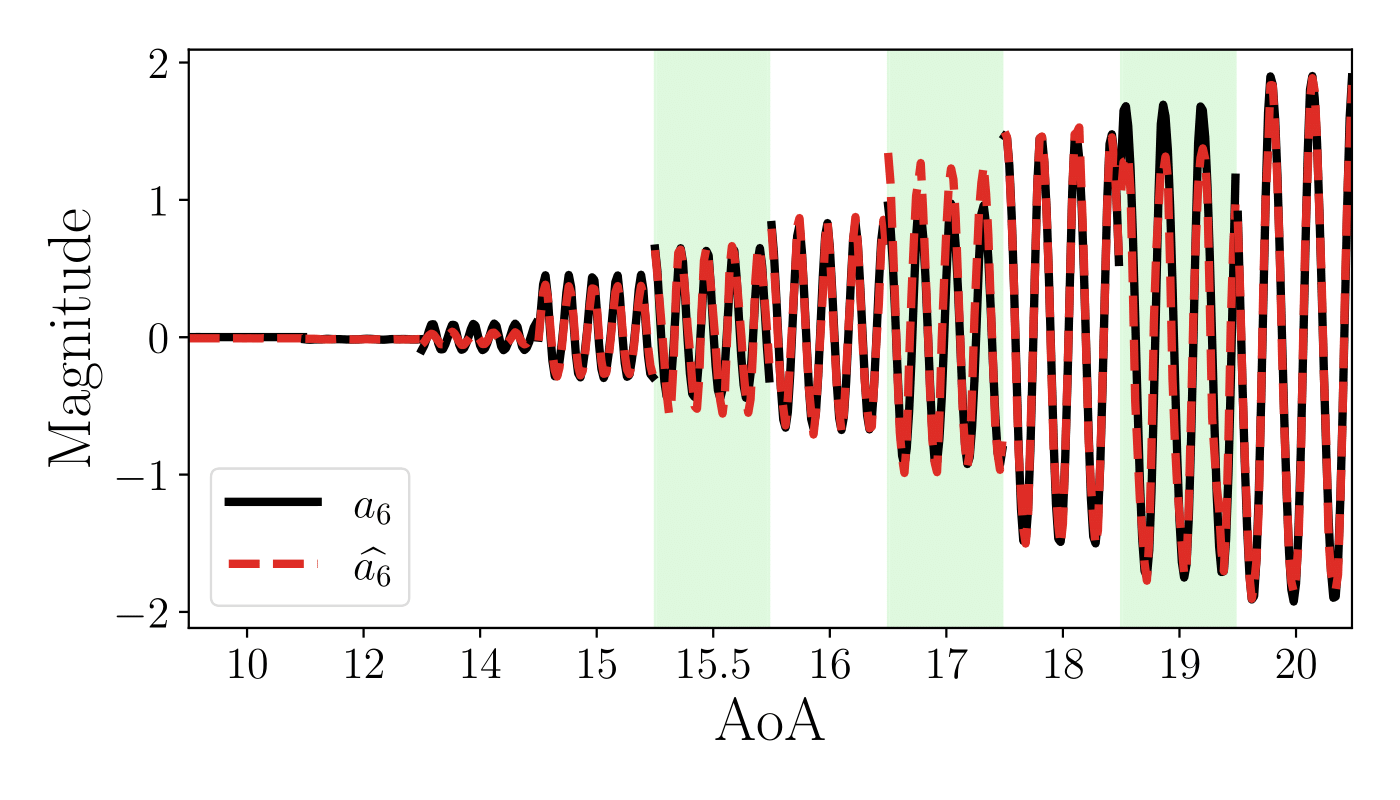}
			\caption{Mode 6}
		\end{subfigure}%
		\caption{Comparison between the true POD amplitudes $a_i$ (black lines) and the model estimates  
			$\widehat{a}_i$ (SNN-POD estimation red dashed lines) on the test dataset, with $i = 1,2,3,4,5,6$ (a,b,c,d,e,f respectively). 
			The test snapshots are organized into chronological time and sorted by angles of attack-angles of attack seen during training are displayed on white background, those not seen are highlighted with light green backgrounds} 
		\label{fig:POD_coef_est}
	\end{figure}

	A more direct evaluation of the model performance is provided by figure \ref{fig:POD_coef_est}, which compares the estimated POD coefficients with their true values at all angles.
	Again, snapshots have been organized by increasing angle of attack and in chronological order so that the dynamics of the reconstruction can be evaluated, although no time dependence is accounted in the model.
	Interpolated angles are indicated by green shaded regions.  
	As noted in Fig. \ref{fig:U_POD_modes}, strong non-monotonic variations are observed in the amplitudes of the first three modes around AoA $=14 ^\circ$, while the evolution appears to be more gradual for AoA $\ge 15 ^\circ$.
	The time-averaged values of the amplitudes appear to be correctly predicted for all angles of attack, with the largest discrepancy observed for mode 3
	at AoA $=15.5 ^\circ$ (note that the true value does not correspond to a direct interpolation  of the angles AoA $=15^\circ$ and AoA $=16^\circ$, a trend which
	the model captures but overpredicts).  
	
\begin{figure}[h]
		\centering
		\includegraphics[width=0.85\linewidth]{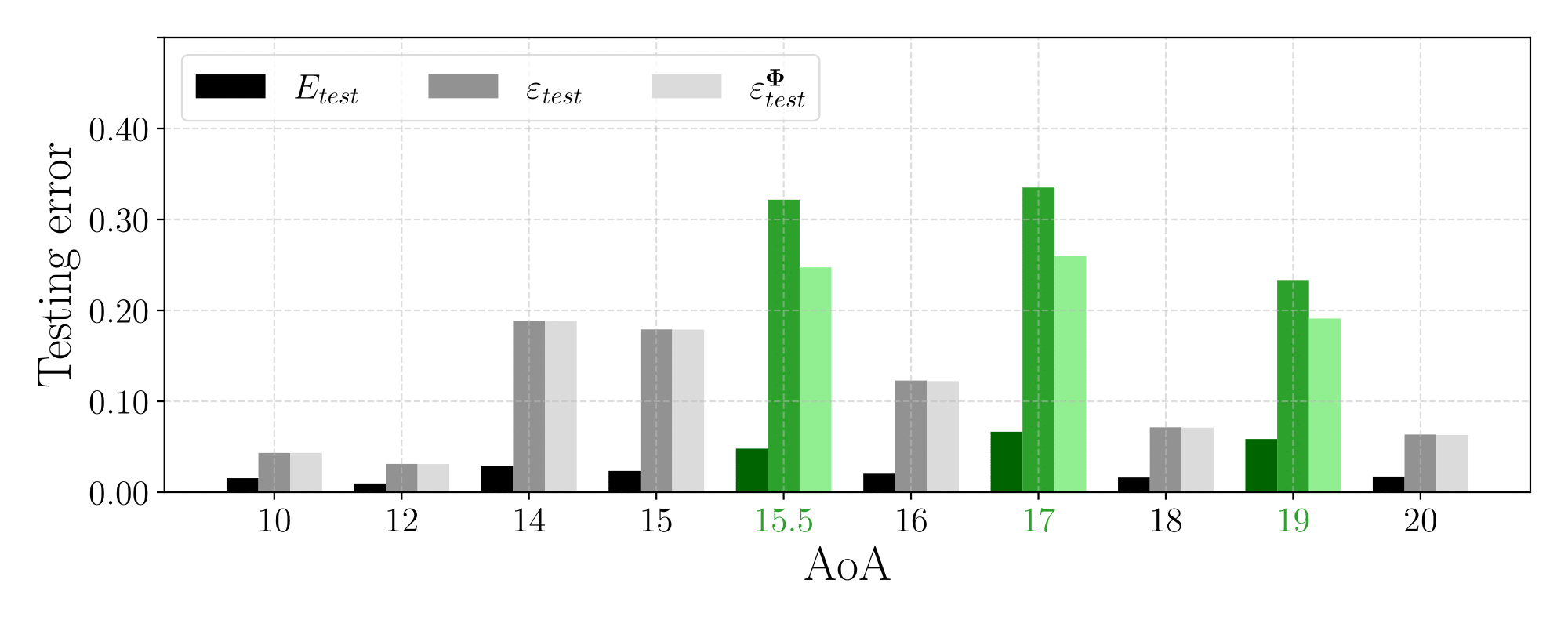}
		\caption{Testing mean error $E_{test}$, fluctuating error $\varepsilon_{test}$ and projected fluctuating error $\varepsilon_{test}^{\mathbf{\Phi}}$ on testing datasets for each AoA configuration. The three interpolated AoA $\{15.5^\circ;17^\circ;19^\circ\}$ are highlighted in green}
		\label{fig:reconstruction_errors}
	\end{figure}

	The mean, fluctuation and projected fluctuation errors defined in section \ref{sec:SNN_architecture} are computed individually for each angle of attack of the test dataset and shown in Fig. \ref{fig:reconstruction_errors}.
	As expected, errors are higher for interpolated AoA than for sampled AoA.
	Nonetheless, the mean reconstruction error remains below $6.6\%$ (maximum observed for $17^\circ$), while the largest fluctuation errors $\varepsilon_{\text{test}}$ are about $30\%$ for $15.5^\circ$ and $17^\circ$, and reach about $23\%$ for $19^\circ$. 
	The projection error is essentially equal to the fluctuation error for sampled angles but about 25\% lower for interpolated angles, a fraction that tends to decrease as the AoA increases ($25\%$ for $15.5^\circ$ and $17^\circ$, and under $20\%$ for $19^\circ$).
	Overall, this points to a moderate influence of the POD basis on the model error.
	It can be noted that the projection error is largest for $15.5^\circ$, despite the finer AoA resolution in that range. 
	This suggests that significant changes occur in this range of angle of attack, which is also consistent with the higher errors observed at the angles $14^\circ$, $15^\circ$, and to a lesser extent $16^\circ$.  
	This point will be further discussed in Sec. \ref{sec:specific_model}.

	
	\subsection{Robustness of the model}
	\label{sec:robustness_analysis}
	\subsubsection{Evaluation of the sensor placement strategy}
	\label{sec:sensor_location_sensitivity}
	
\begin{figure}[h]
		\centering
		\includegraphics[width=0.5\linewidth]{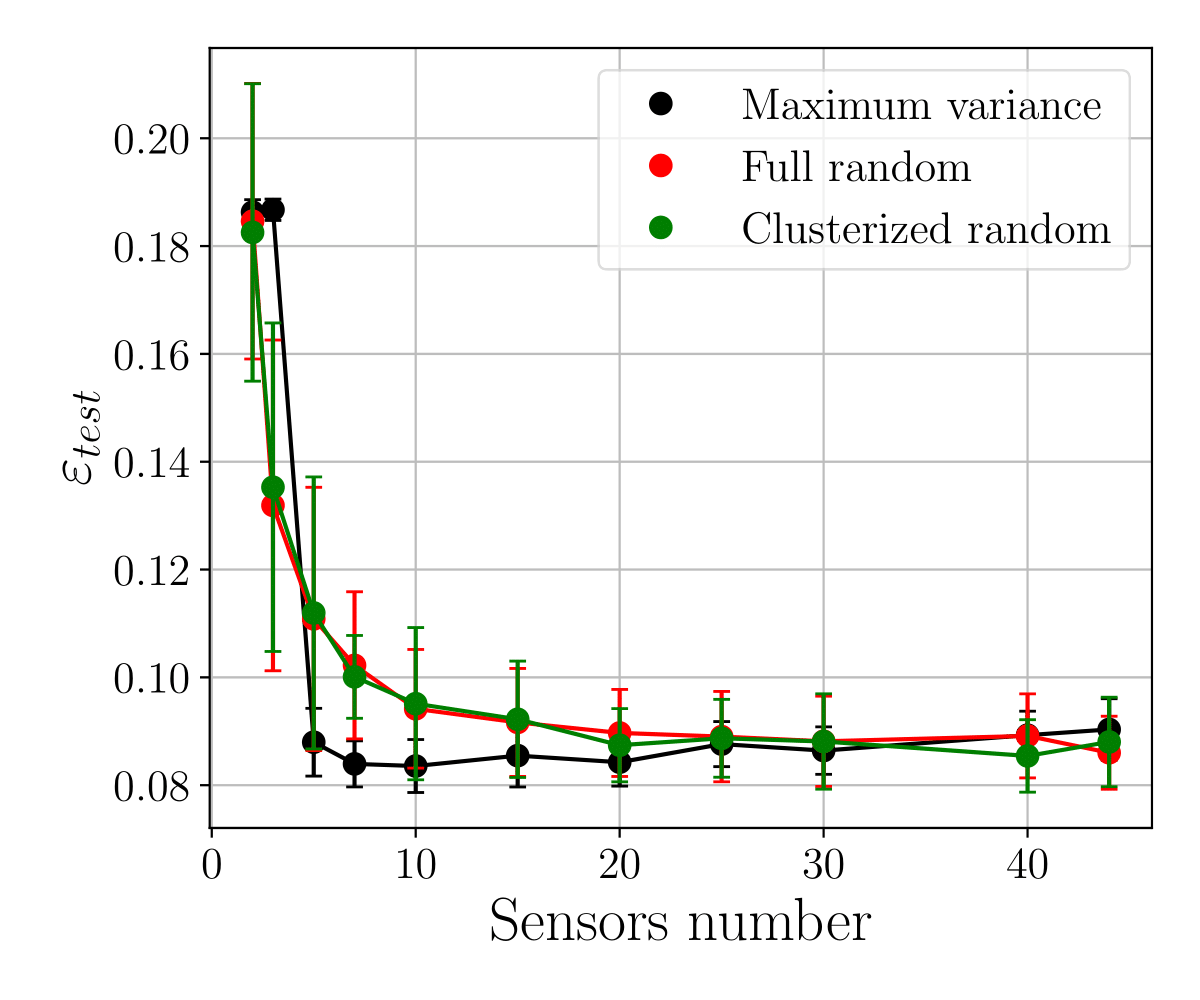}
		\caption{Sensitivity study of the pressure sensors number}
		\label{fig:sensors_number_sensitivity}
	\end{figure}

	In order to assess the influence of sensor placement, a comparative analysis involving three sensor selection approaches was carried out: 
	(i) the variance-maximization strategy described in Sec. \ref{sec:sensors_placement_description}
	(ii) a fully random placement where all sensors are placed randomly, 
	(iii) a clustered random placement where preliminary clustering is first 
	performed and then one sensor location is drawn with a uniform probability from each cluster, which ensures a more uniform sensor distribution than the full random approach.  
	Evaluation on the testing set defined above was performed over 30 realizations of the model, which were trained on the same training set with different initial conditions.
	In all cases a minimum distance $d_{min}$ was imposed between sensors ($d_{min} = c/100 = $ \num{1.25e-2} m as used in the maximum variance approach). 
	Fig. \ref{fig:sensors_number_sensitivity} shows the evolution of the fluctuation error with the number of sensors for the three approaches.
	Unsurprisingly, as the number of sensors $p$ increase, the reconstruction error decreases towards a value that is the same for all strategies (within statistical uncertainty).
	However, when the number of sensors is small ($3 \le p \le 15$),  the reconstruction error is lower for the maximum variance strategy, with significantly less variability. This validates the choice of $p=7$ sensors made in the study.
	
	\subsubsection{Modal splitting sensitivity study}
	\label{sec:modal_splitting_robustness}
	
	In this section the robustness of the model with respect to the splitting strategy is evaluated.
	Different architectures based on the shallow neural network framework were compared in terms of performance and of training cost. 
	For each architecture, 30 models were trained and evaluated, and average values are reported in Table \ref{tab:cpu_cost}.
	
	The following models were considered : 
	\begin{itemize}
		\item the SNN-POD model described  Sec. \ref{sec:methodology}, based on two independent networks for the steady and unsteady component of the POD amplitudes
		\item the SNN model which directly learns the mapping between the pressure measurements and the velocity field (no latent space reduction)
		\item the SNN-POD-0 model which learns the mapping in POD space with a single network (all POD amplitudes are mixed together)
		\item the SNN-POD-50 model which learns the mapping in POD space with a network for each POD mode.
	\end{itemize}
	
	The SNN has both a high training cost and a high reconstruction error, which illustrates the importance of learning latent features. 
	The SNN-POD-50 has the lowest reconstruction error but also the highest training cost, which is due to the high number of networks.
	Unsurprisingly, the lowest training cost is observed for the SNN-POD-0, but it also has the highest reconstruction error. 
	
	
	\begin{table}[h]
		\centering
		\begin{tabular}{ccccc}
			\hline
			\hline
			& SNN-POD & SNN & SNN-POD-0 &  SNN-POD-50 \\
			\hline
			Number of SNNs trained & 2 & 1 & 1 &  50 \\
			\hline
			$E_{test}$ & \num{2.0e-2} & \num{3.2e-2} & \num{3.9e-2} &  \num{1.9e-2} \\
			\hline
			$\varepsilon_{test}$ & \num{8.5e-2} & \num{1.4e-1} & \num{1.6e-1}  & \num{8.0e-2} \\
			\hline
			CPU training time (s) & 330 & 3530 & 220 &  6000 \\
		\end{tabular}
		\caption{Average training cost and MSE for different SNN-based architectures. See text for details}
		\label{tab:cpu_cost}
	\end{table}
	
\begin{figure}[h]
		\centering
		\includegraphics[width=0.8\linewidth]{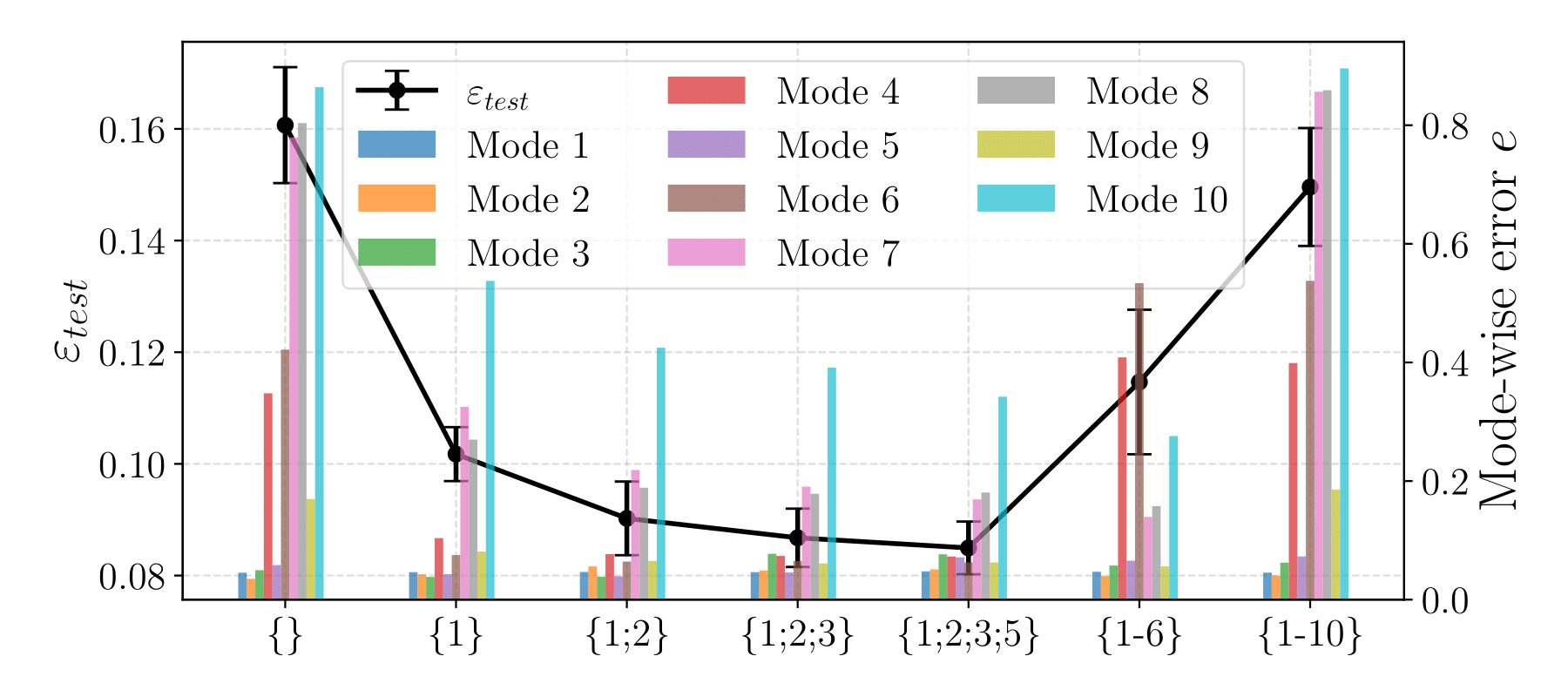}
		\caption{Reconstruction error $\varepsilon_{test}$ and mode-wise reconstruction errors $e$ of the first $10$ POD modes for different scale separations. Statistics are computed over 30 independent training runs}
		\label{fig:steady_modes_robustness}
	\end{figure}

	The sensitivity of the model to the modal splitting was next examined.  
	Different choices for the two-mode split were evaluated, with comparison to the no-split case SNN-POD-0, denoted as $\{\}$ in Fig. \ref{fig:steady_modes_robustness}.
	Six mode-splitting configurations were investigated, where one to ten modes are selected to be reconstructed by one shallow neural network, while the remainder of the 50 POD modes is estimated by another. 
	In the first four configurations, ($\{1\}$, $\{1;2\}$, $\{1;2;3\}$, $\{1;2;3;5\}$), the output of the first network consisted of an increasing number of modes that are all considered steady according to criterion Eq.\eqref{eq:mode_criterion}.
	In contrast, for the other two configurations $\{1-6\}$ and 
	$\{1-10\}$, the output of the first network respectively consisted of 
	the first six and ten most energetic modes (energy-based splitting). This means that the first network reconstructs both steady and unsteady modes.

	Fig. \ref{fig:steady_modes_robustness} represents the fluctuation error for each configuration on the portion of the test dataset described in section \ref{sec:results} limited to sampled angles only.  
	It is observed that the strategy used to separate steady and unsteady modes has a significant impact on the global reconstruction error, with a drop
	of nearly 40\% if at least one steady mode is trained independently from the others.
	For these configurations, the error varies between  $8.5\%$ and $9.1\%$, so that it depends only weakly on the exact splitting. 
	In contrast, the performance drops to a level comparable to that of the no-split model (SNN-POD-0, $\{\}$) when training $\text{SNN}_{st}$ on unsteady modes, as it is the case for $\{1-6\}$ and $\{1-10\}$.
	These results illustrate both the importance and the robustness of the mode-splitting strategy.
	
	Fig. \ref{fig:steady_modes_robustness} also shows the mode reconstruction error
	for each of the first ten POD modes, where the reconstruction error $e_i$ for the $i$-th POD mode is defined as : 
	\begin{equation}
		e_i = \frac{\left\Vert a_i - \widehat{a}_i \right\Vert_2}{\left\Vert a_i \right\Vert_2}
	\end{equation}
	Examination of the errors $e_i$ reveals that the steady modes (1,2,3,5) are well recovered in all cases.
	In view of the loss formulation in Eq. \eqref{eq:loss}, it is expected that the error increases with the POD mode index.
	Including mode 4 (which is unsteady, as can be seen in figure \ref{fig:U_POD_modes}) with the steady modes significantly degrades its reconstruction, particularly evident in the $\{1-6\}$ configuration where the first six modes are learnt together.
	This further confirms the relevance of the steadiness-based splitting.
	
	\subsection{Combining SNN-POD models}
	\label{sec:specific_model}
	
	As reported in section \ref{sec:evalsnnpod}, the model globally trained over the full range of angles of attack is generally able to recover the time-averaged features of the flow such as the wake deflection angle (and therefore the angle of attack), while flow reconstruction of the time-dependent dynamics can be more challenging, particularly at the onset of separation where fluctuations are relatively small.  
	A solution to improve the reconstruction could therefore be to combine the global model with local models that would be trained for specific flow regimes. 
	The global model would first be used to infer the steady part of the flow and in particular the configuration angle of attack, based on which a model specifically trained for this flow regime could be selected. 
	
	\begin{center}
		\begin{table}[h!]
			\centering
			\begin{tabular}{ l }
				\hline
				\hline
				\textbf{ Combined model (SNN-POD-C) } \\
				1: from pressure input evaluate  $\mathbf{\widehat{a}_{st}}$ with global model (steady global network)  \\
				2: compute steady flow field $\mathbf{\widehat{u}_{st}} = \mathbf{\widehat{a}_{st}} \mathbf{\Phi_{st}}$ \\
				3: compute deflection angle $\alpha$ for each reconstructed steady flow field \\
				4: \textbf{if} $\alpha (x/c = 4) \in \left[ \alpha_1 ; \alpha_2 \right]$ (or other suitable criterion) \\
				5: \hspace{0.5cm} evaluate $\mathbf{\widehat{a}_{st}^l}$ and $\mathbf{\widehat{a}_{un}^l}$ with local model  \\
				6: \hspace{0.5cm}  reconstruct $\widehat{\mathbf{u}} = \mathbf{\langle u \rangle^{l}} + \mathbf{S_{st}^l} \mathbf{\widehat{a}_{st}^l} \mathbf{\Phi_{st}^{l}} + \mathbf{S_{un}^l} \mathbf{\widehat{a}_{un}^l} \mathbf{\Phi_{un}^l}$  \\
				7: \textbf{else} evaluate $\mathbf{\widehat{a}_{un}}$ with global model (unsteady global network) \\
				8: \hspace{0.5cm}  reconstruct $\widehat{\mathbf{u}} = \mathbf{\langle u \rangle} + \mathbf{S_{st}} \mathbf{\widehat{a}_{st}} \mathbf{\Phi_{st}} + \mathbf{S_{un}} \mathbf{\widehat{a}_{un}} \mathbf{\Phi_{un}}$  \\
				8 : \textbf{end if} \\
				\\
				\hline
			\end{tabular}
			\caption{Combined model SNN-POD-C based on global and local models} 
			\label{tab:specific_model}
		\end{table}
	\end{center}
	
	To test this idea, a local model was developed
	by training a SNN-POD model on the angles $14^\circ,15^\circ,16^\circ$.
	It is emphasized that the POD basis and mean field are generally different for the global and local models since they are associated with different training sets - the suffix $\mathbf{l}$ will be used for the local model.   
	The approach is summarized in Tab. \ref{tab:specific_model}.
	The combination of the global and the local models, referred to as SNN-POD-C, was then evaluated for the testing dataset in the range $[14^\circ,16^\circ]$.

	\begin{figure}[h]
		\begin{subfigure}{\textwidth}
			\centering
			\includegraphics[width=\linewidth]{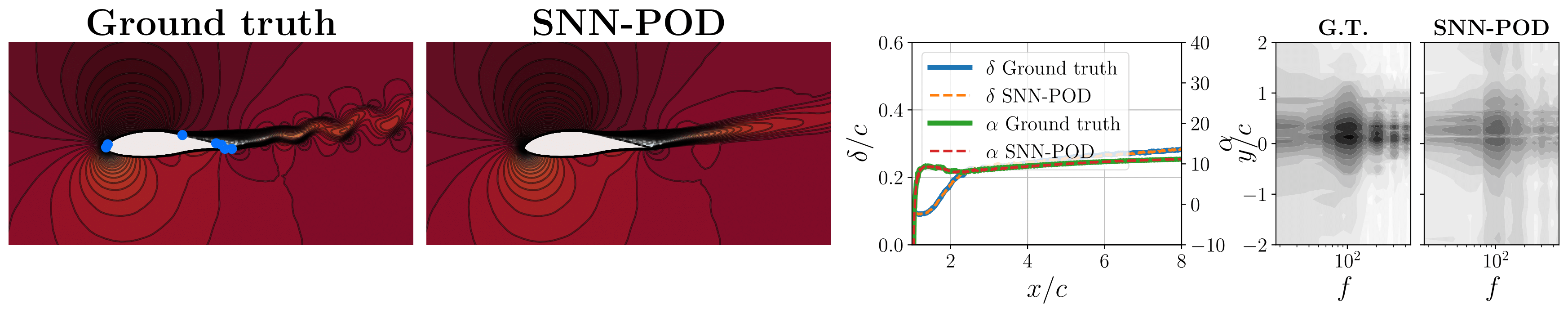}
			\caption{AoA = $14^\circ$, global model}
		\end{subfigure}
		\\
		\begin{subfigure}{\textwidth}
			\centering
			\includegraphics[width=\linewidth]{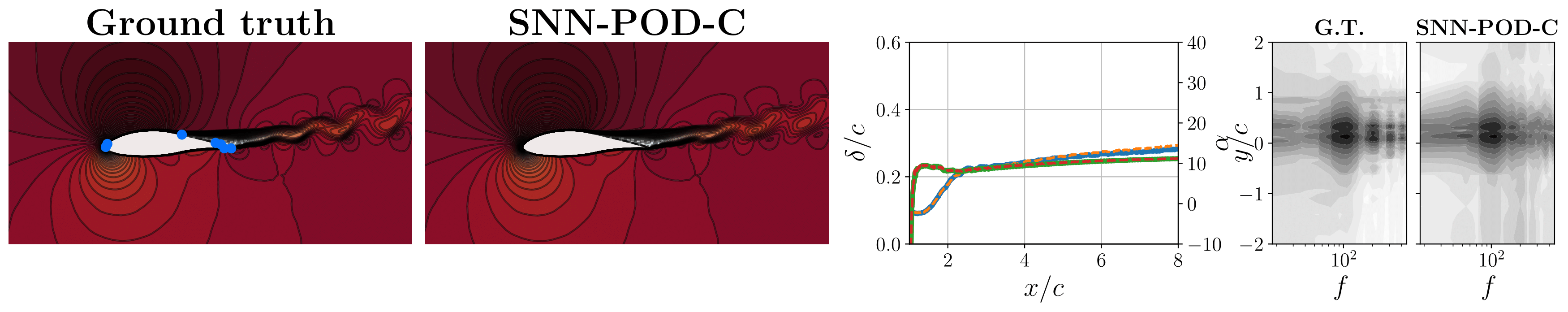}
			\caption{AoA = $14^\circ$, combined model}
		\end{subfigure}
		\caption{Side-by-side comparison of streamwise velocity $u(\mathbf{x},t_s)$ contours at snapshot $t_s=150$s of the ground truth obtained from URANS (first plot from left to right) and its SNN-POD / SNN-POD-C estimation (second plot). The wake widths and wake deflection angles comparison between the ground truth and the estimation are plotted on the third subfigure. The fourth and fifth plots correspond to the spectrograms of the pre-multiplied velocity spectra $f S_u$ of the ground truth (G.T., left) and estimation (SNN-POD / SNN-POD-C, right) at the $x/c = 2$ section. The same logarithmic colormap is applied for both spectrograms}
		\label{fig:SNN_POD_evaluation_AoA14_specific}
	\end{figure}
	
	Figure \ref{fig:SNN_POD_evaluation_AoA14_specific} compares the SNN-POD and SNN-POD-C reconstructions with the ground truth  
	at the challenging angle of AoA $=14^\circ$, corresponding to the onset of separation. 
	Comparing the  two left columns on each row shows that the instantaneous features of the wake are more accurately captured by the combined model.
	The mean characteristics of the wake (width and deflection angle), shown in the middle column, are equally well recovered by both models. 
	Comparison of the spectrograms (two right-most columns) confirms that the SNN-POD-C model provides a more accurate estimate of the fluctuation level, which is underestimated by the SNN-POD model
	
\begin{figure}[h]
		\centering
		\includegraphics[width=0.75\linewidth]{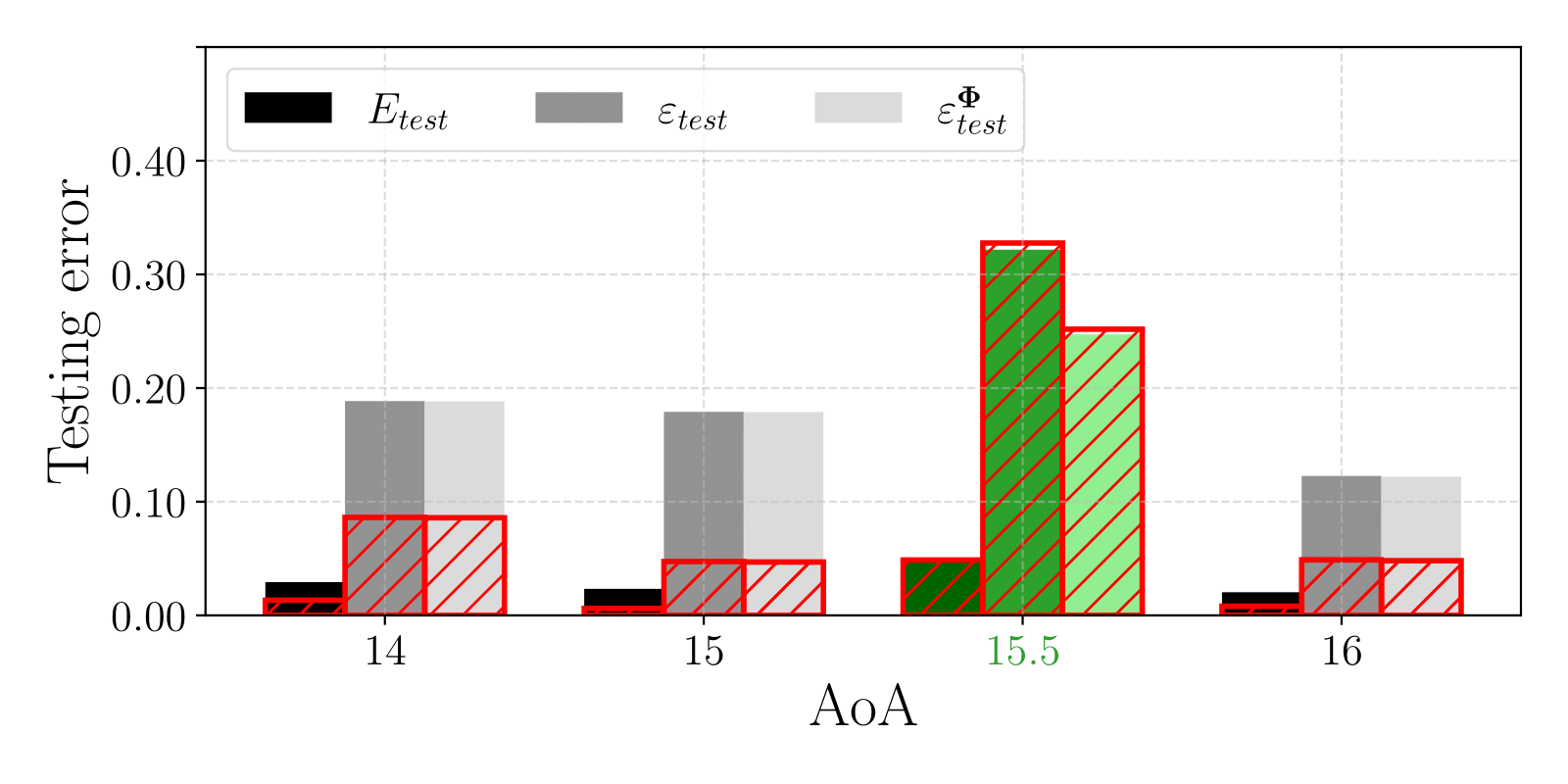}
		\caption{Mean error $E_{test}$, fluctuating error $\varepsilon_{test}$ and projected fluctuating error $\varepsilon_{test}^{\mathbf{\Phi}}$ for transitional AoA without (colored bars) and with (red hatched bars) the use of a local model. The interpolated AoA $=15.5^\circ$ is highlighted in green}
		\label{fig:reconstruction_errors_specific}
	\end{figure}

	Fig. \ref{fig:reconstruction_errors_specific} presents the three types of reconstruction errors for the standard SNN-POD (shaded areas) and the 
	combined SNN-POD-C models (denoted by red hatched bars) for the range $[14^\circ, 16^\circ]$. 
	All three errors are reduced by a factor of 2 for the angles $14^\circ, 15^\circ, 16^\circ$, which are seen during training.
	Interestingly, no reduction is observed for the interpolated angle of $15.5^\circ$, which shows that the gain associated with specific training does not necessarily lead to better generalization properties.

	The main outcome of this coupling framework is that the global model demonstrates good performance in reconstructing instantaneous fields for fully attached and separated regimes, while its training over a wide AoA coverage enables interpolation to unseen configurations. Moreover, thanks to the mode-splitting, the global model can infer the AoA regime, allowing it to be coupled with local models that achieve higher accuracy on the transitional AoA regime.
	
	\section{Conclusion}
	\label{sec:conclusion}
	
	This work describes an efficient data-driven  methodology for the one-shot reconstruction of instantaneous velocity fields around a thick 2D airfoil from sparse wall-pressure measurements. 
	The dataset is generated from 2D high-Reynolds-number URANS simulations over a range of angle of attacks between 10° and 20°, spanning a variety of flow regimes ranging from attached flow to full separation. 
	Proper Orthogonal Decomposition provides a low-dimensional representation of the velocity field that can be split into steady modes, the amplitudes of which depend mostly on the angle of attack, and a remainder of unsteady modes. 
	The model is constituted by two shallow neural networks that independently predict the unsteady and steady parts of the POD amplitudes from limited surface pressure measurements. 
	A sensor placement strategy based on a a maximization variance criterion was found to achieve better accuracy than randomized placements when the number of sensors is kept low. 
	The model is trained on a coarse discretization of the angle of attack range ($1-2^\circ$) in order to test its generalization abilities.
	The maximum reconstruction error( MSE) was found to be 2.9\% for sampled angles and 6.6\% for interpolated angles. 
	In all cases the time-averaged flow features at a given angle of attack were well recovered from a single-time set of instantaneous pressure measurements.
	Reconstruction  of the time-dependent features could be improved by combining the model trained over the full range of angles of attack with one trained over a more local range.  
	It is emphasized that this proposed strategy merely aims to highlight one of the multiple possibilities offered by the SNN-POD framework.
	
	Several perspectives for extending the study can be identified. 
	Firstly, in the present manuscript, only variations in AoA were considered to define different flow configurations; however, it would be relevant to investigate the performance of SNN-POD under changes in other parameters such as turbulence intensity or Reynolds number. 
	Secondly, the current framework is limited to training on 2D URANS simulations, which cannot capture three-dimensional effects, which are particularly important in near flow separation conditions for thick airfoils. 
	A natural extension is thus the application of SNN-POD to higher-fidelity 3D datasets for the reconstruction of fully unsteady 3D flow fields, ultimately making it a valuable tool for experimentalists. 
	Lastly, the reconstruction methodology has so far been assessed only on static simulations with constant AoA, thereby neglecting the additional nonlinearities and hysteresis effects characteristic of dynamic stall. 
	Extending the model to dynamic AoA variations would constitute a next step, enabling the evaluation of the model under more complex and unsteady operating conditions. 
	To this end, more sophisticated approaches may be required, such as 
	incorporating temporal history to capture coherent turbulent structures within the model.

		\paragraph{Acknowledgments}
		This work was granted access to the HPC resources of IDRIS under the allocation 2025-AD012A16542 made by GENCI.

		\paragraph{Funding Statement}
		This research was performed within the French-Swiss project MISTERY funded by the French National Research Agency (ANR PRCI grant no. 266157) and the Swiss National Science Foundation (grant no. 200021L 21271). The authors are thankful to the anonymous reviewers for their helpful suggestions.
		
		\paragraph{Competing Interests}
		The authors declare no competing interests exist.
		
		
		\paragraph{Author Contributions}
		Conceptualization: Q.B; B.P; C.B; E.G. Methodology: Q.B; B.P. Data curation: Q.B; C.B; E.G. Data visualization: Q.B. Writing - original draft: Q.B. Writing - review \& editing: Q.B; B.P; C.B; E.G. All authors approved the final submitted draft.
		
		\paragraph{Supplementary Material}
		State whether any supplementary material intended for publication has been provided with the submission.
		
		\bibliographystyle{apalike}
		
		\bibliography{2D_Flow_reconstruction_paper_bp}

\end{document}